\documentclass[12pt]{article}

\def \real{{\rm I\!R}} 
\def \GI{\mathrm {GI}}

\def \argmax{\mathrm{argmax~}} 

\def \sumi{\sum_{i=1}^I}
\def \sumj{\sum_{j=1}^J}

\def \pij{\hat p_{ij}}
\def \pi{\hat p_{i.}}
\def \pj{\hat p_{.j}}

\usepackage{amsmath}
\usepackage{graphicx}
\usepackage{setspace}
\usepackage{subcaption}
\usepackage{epstopdf}
\usepackage{cuted}
\usepackage{enumitem}
\usepackage{color}

\usepackage{lineno,hyperref}
\usepackage{graphicx}
\usepackage[latin1]{inputenc}
\usepackage{amsmath,amsfonts,graphics,epsfig,calc,multirow}
\usepackage{amssymb,natbib,rotating}
\usepackage{epsf}
\usepackage{setspace}

\def \real{\mathbb R}

\newtheorem{theorem}{Theorem}

\begin{document}
\title{Parallel Coordinate Order for High-Dimensional Data}
\author{Shaima Tilouche, Vahid~Partovi~Nia\footnote{Ecole Polytechnique de Montreal, 2900 Edouard Montpetit, Montreal, Quebec, Canada. Corresponding author \texttt{vahid.partovinia@polymtl.ca} 
}, and Samuel~Bassetto  } 
\date{\today}

\maketitle {\narrower\smallskip\centerline{\bf Abstract} \noindent  
Visualization of high-dimensional data is counter-intuitive using conventional graphs. Parallel coordinates are proposed as an alternative to explore multivariate data more effectively. However, it is difficult to extract relevant information through the parallel coordinates when the data are high-dimensional with thousands of lines overlapping. 
The order of the axes determines the perception of information on parallel coordinates. 
Thus, the information between attributes remain hidden if coordinates are improperly ordered. Here we propose a general framework to reorder the coordinates. This framework is general to cover a large range of data visualization objective. It is also flexible to contain many conventional ordering measures. Consequently, we present the coordinate ordering binary optimization problem and enhance towards a computationally efficient greedy approach that suites high-dimensional data. Our approach is applied on wine data and on genetic data. The purpose of dimension reordering of wine data is highlighting attributes dependence. Genetic data are reordered to enhance cluster detection. The presented framework shows that it is able to adapt the measures and criteria tested.
}


\maketitle

\section{Introduction}
\label{sec:Intro}
When data are high-dimensional, representing each attribute marginally may lead to an incomplete or unclear visualization. 
Multidimensional graphs such as scatter plot matrices, glyphs, and parallel coordinates are proposed to facilitate multivariate data exploration. Here we focus on parallel coordinates which \citet{Ocagne:ParallelBook:1885}
 invented, primarily as a two-dimensional diagram to approximate the graphical computation of a mathematical function using \emph{nomogram}. Parallel coordinates are further studied by  \citet{Inselberg1985} to allow the visualization of multidimensional data on a transformed two-dimensional space.

Suppose the data matrix contains $n$ observations in rows and $p$ attributes in columns. A common data visualization representation is scatter plot of data in orthogonal coordinates, where each axis is an attribute and each observation is a point. This representation is limited to maximum $p=3$ attributes. In parallel coordinates representation, axes are parallel lines and each observation is a line, passing through each coordinate \citep{albazzaz2006historical}. This technique extends  data visualization for $p > 3$. 

Several parallel coordinates software have been developed so far. Some of them like \textit{XDAT} and \textit{XMDVTool} are interactive and some others like \textit{Statistica} and \emph{ggparallel} R package are not. Software visualization tools mostly provide options such as applying filters, data clustering, and switching coordinates for a better visualization. Theoretically, there is no limit on the number of observations or the number of attributes. However, when the number of observations is large, many lines overwhelm the display, and the parallel coordinate graph becomes dense to analyze visually. On the other hand, high-dimensional data contain large number of attributes, leading to a wide and an unclear representation.

 \begin{figure}[t]
\begin{subfigure}{0.45\textwidth}
\includegraphics[width=\textwidth, angle=0]{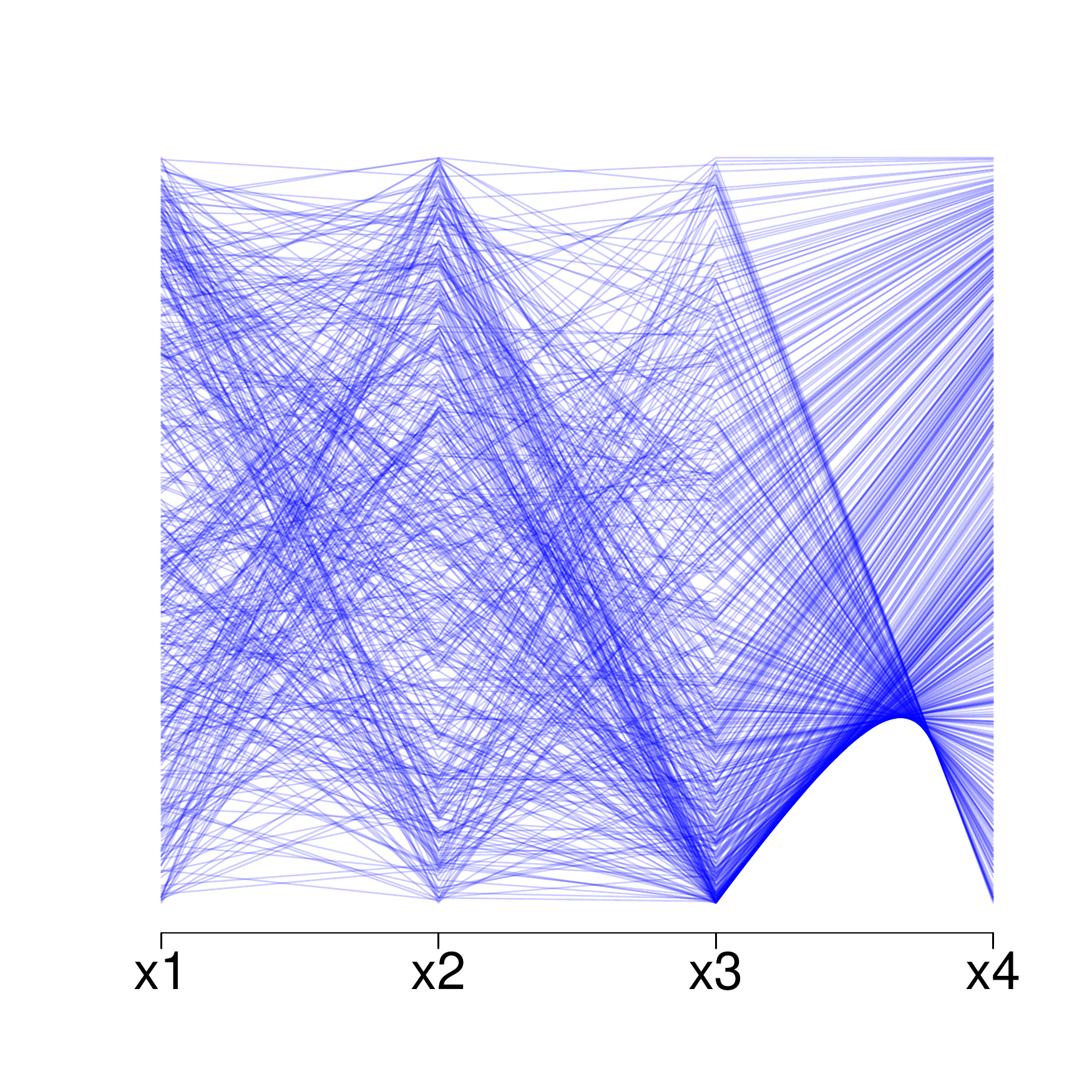}
\caption {~}
\label{fig:withoutorder}
\end{subfigure} 
\begin{subfigure}{0.45\textwidth}
\includegraphics[width=\textwidth, angle=0]{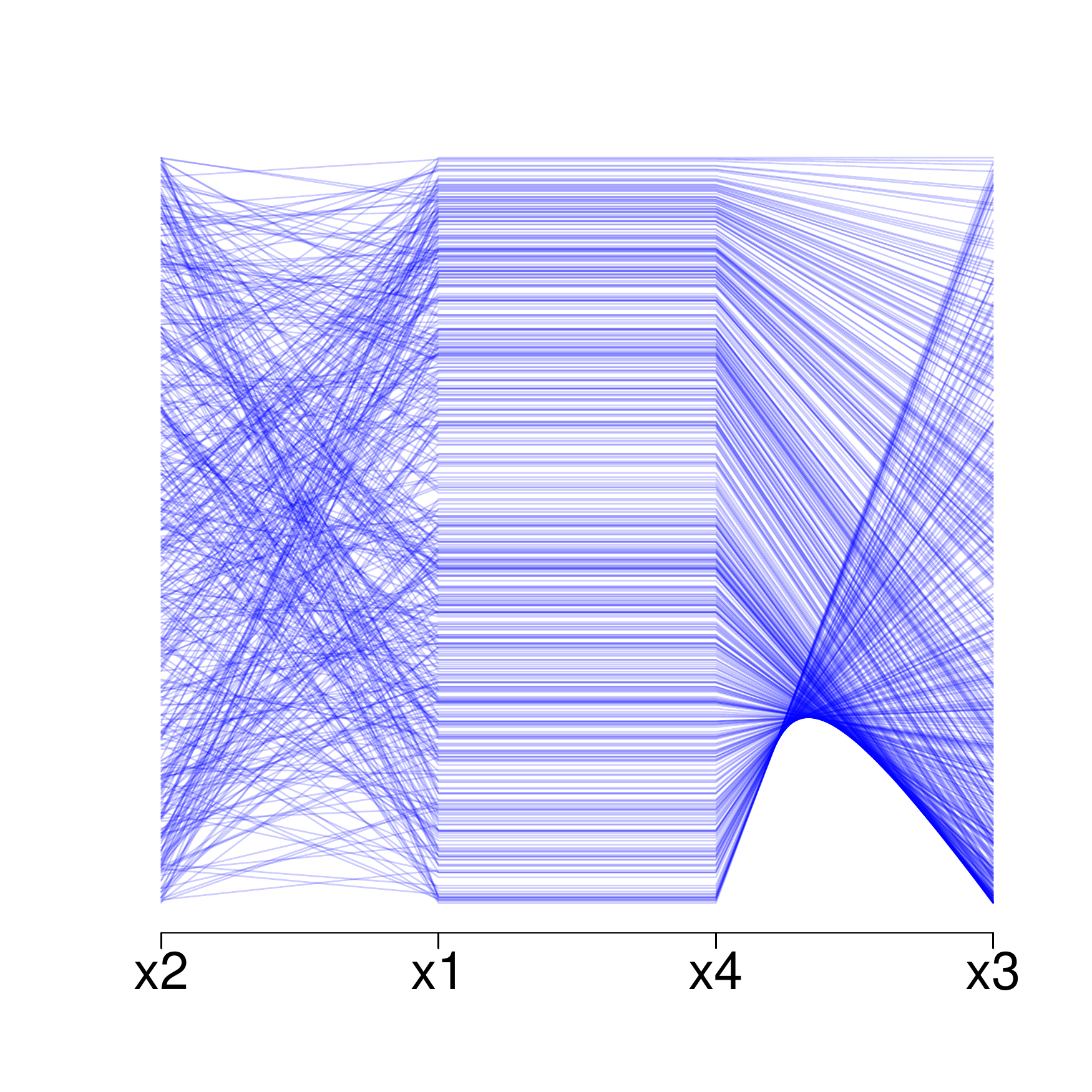}
\caption{~}
\label{fig:ordercorrelations}
\end{subfigure} 
\caption{Parallel coordinate graph when the axes are improperly ordered (left panel), versus properly ordered (right panel) to explore attribute dependence.}
\label{fig:clutterorder}
\end {figure}

 \begin{figure}[h]
\begin{subfigure}{0.45\textwidth}
\includegraphics[width=\textwidth, angle=0]{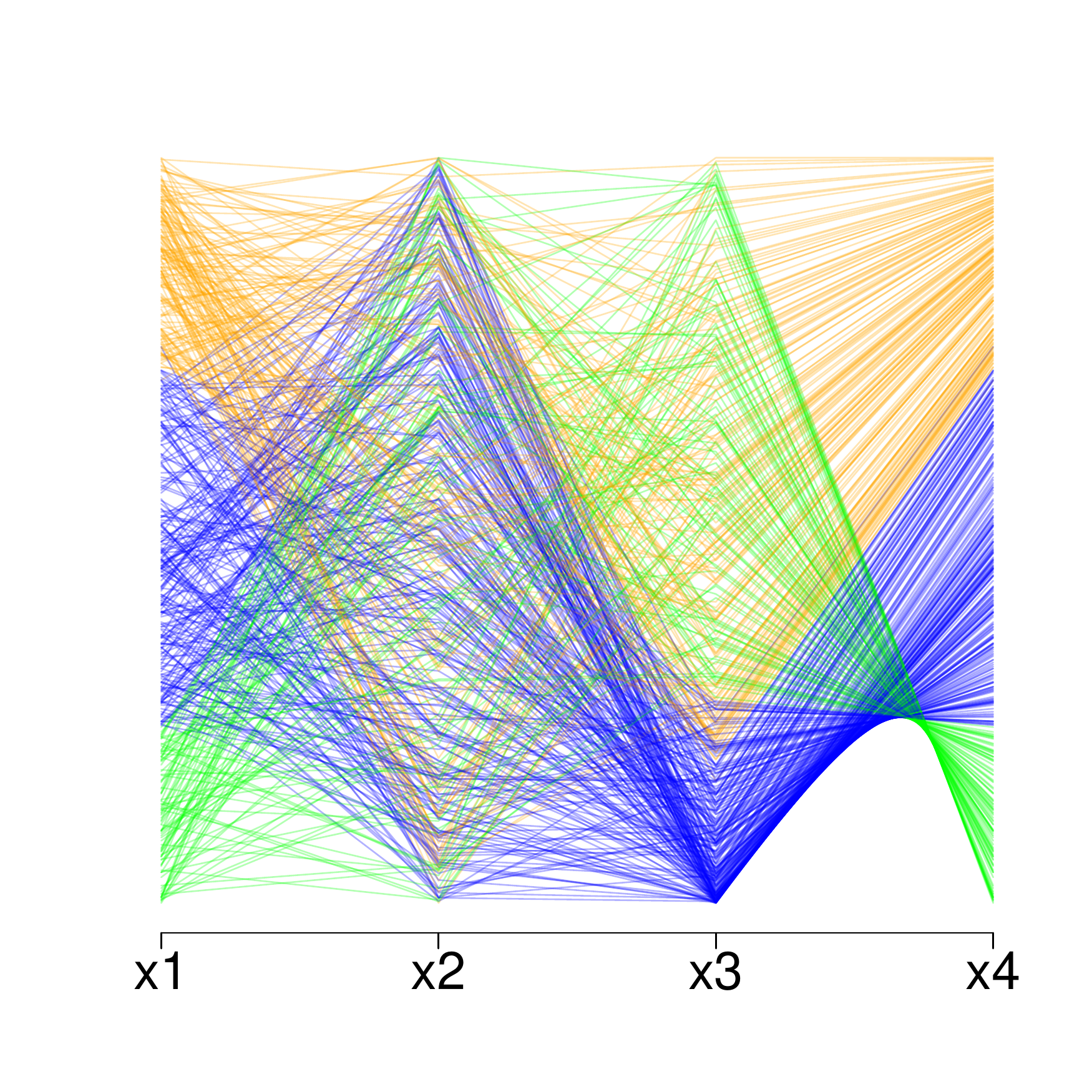}
\caption {~}
\label{fig:noorder}
\end{subfigure} 
\begin{subfigure}{0.45\textwidth}
\includegraphics[width=\textwidth, angle=0]{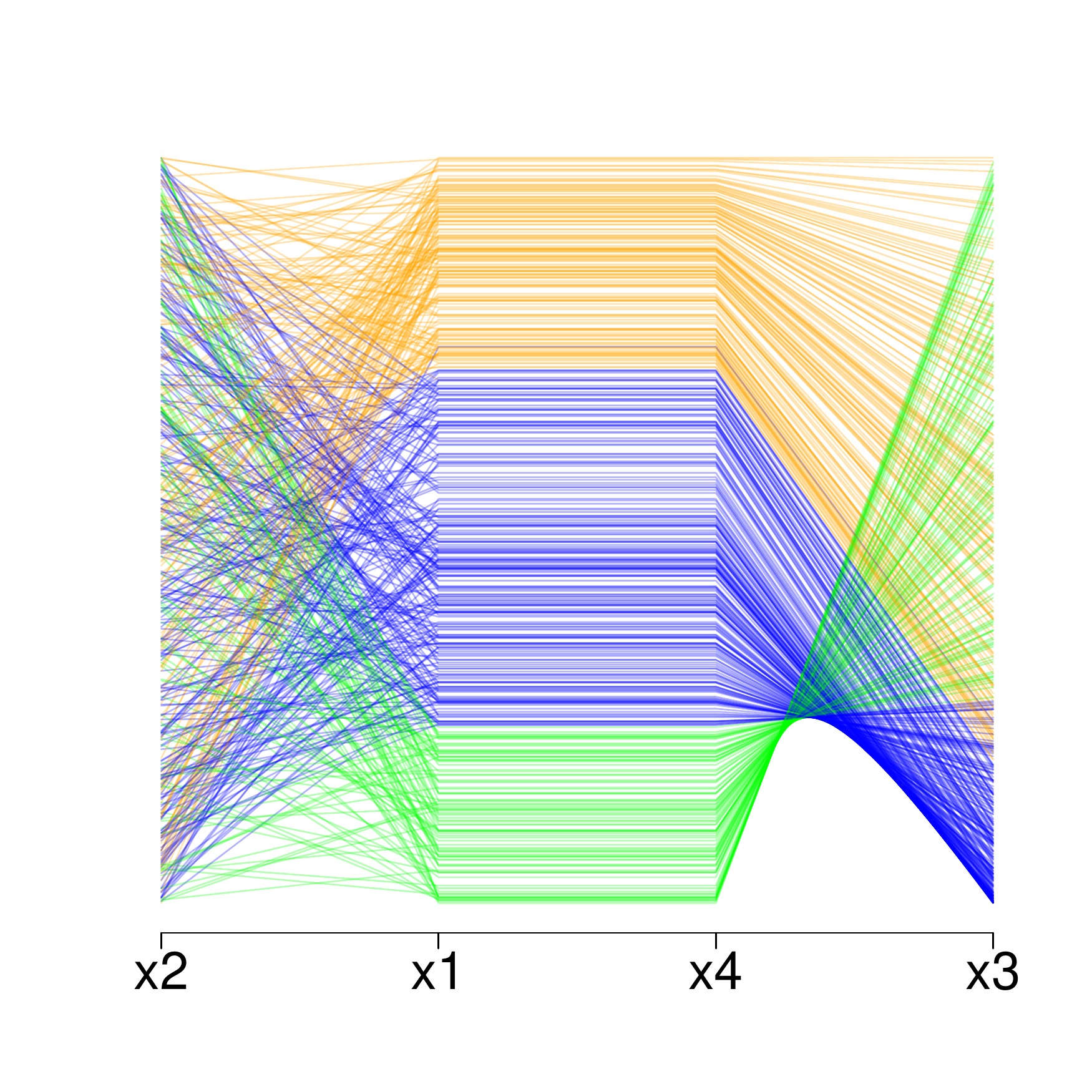}
\caption{~}
\label{fig:ordersep}
\end{subfigure} 
\caption{Parallel coordinate graph when the axes are improperly ordered (left panel) versus properly ordered to explore data separation (right panel).}
\label{fig:seporder}
\end {figure}

Figure~\ref{fig:clutterorder} shows the impact of data reordering on dependence visualization, even on small number of attributes. The left panel shows only the relation between $x_3$ and $x_4$. With a coordinate reordering, two relationships appear, one between $x_1$ and $x_4$ and another between $x_3$ and $x_4$. 
Figure~\ref{fig:seporder} shows that dimension reordering enhances cluster detection. In the left panel, data are separable only by $x_3$ and $x_4$. However, with a proper reordering, the same data are separable by $x_1$ and $x_4$ as well.

Several techniques are proposed to improve the visual exploration of data in parallel coordinates. These techniques aim to reorder attributes, so that data exploration becomes more straightforward. These techniques aim to highlight relations between attributes and to reduce data clutter. They are based on defining a specific criterion. To our knowledge, there is no general framework presented that can contain different purposes of dimension reordering. Our framework attempts to introduce a measure that adapts to the purpose of parallel coordinate visualization. The bivariate measure between each pair of attributes is defined by $2$ probability functions, $F$ and $H$ defining the measuring concept and a third function $G()$ that defines the statistic. 

Our technique is flexible and can be adapted for other purposes like outlier detection, classification, nonlinear correlation, etc. If the purpose of visualization is exploring the linear dependence between attributes, the criterion can be tailored to mimic correlation. If data clustering is of interest, the criterion is tuned to measure data separation. 
For the test part, we mainly focus on two purposes, exploring the dependence between attributes, and exploring data clustering. Two dataset were used, wine dataset, which is commonly used for this kind of problem and a genetic dataset to show the performance of the technique with high-dimensional data.
The achieved tests show that changing the statistic has an impact on the order of attributes and changing the probability functions change the highlighted concept.

The order of coordinates has a visible impact on dependence visualization and on cluster detection. The parallel coordinate display visualizes the inter-coordinate dependence between neighboring dimensions, but does not reveal the dependence between non-adjacent coordinates \citep{peng2004cluttercluster}.

\begin{figure*}[t]
\centering
\begin{subfigure}{0.24\textwidth}
\centering
\includegraphics[width=\textwidth, angle=0]{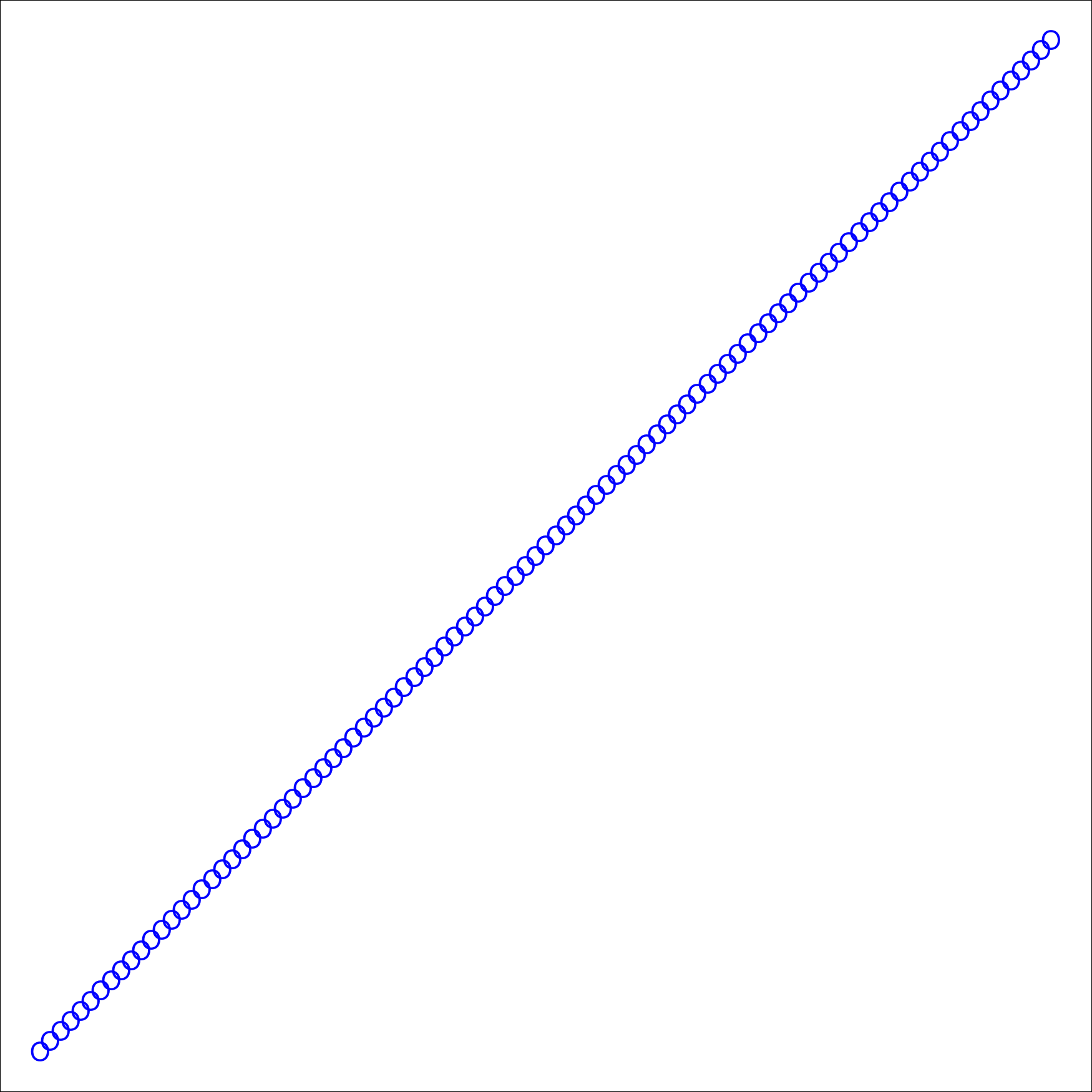}\\
\includegraphics[width=\textwidth, angle=0]{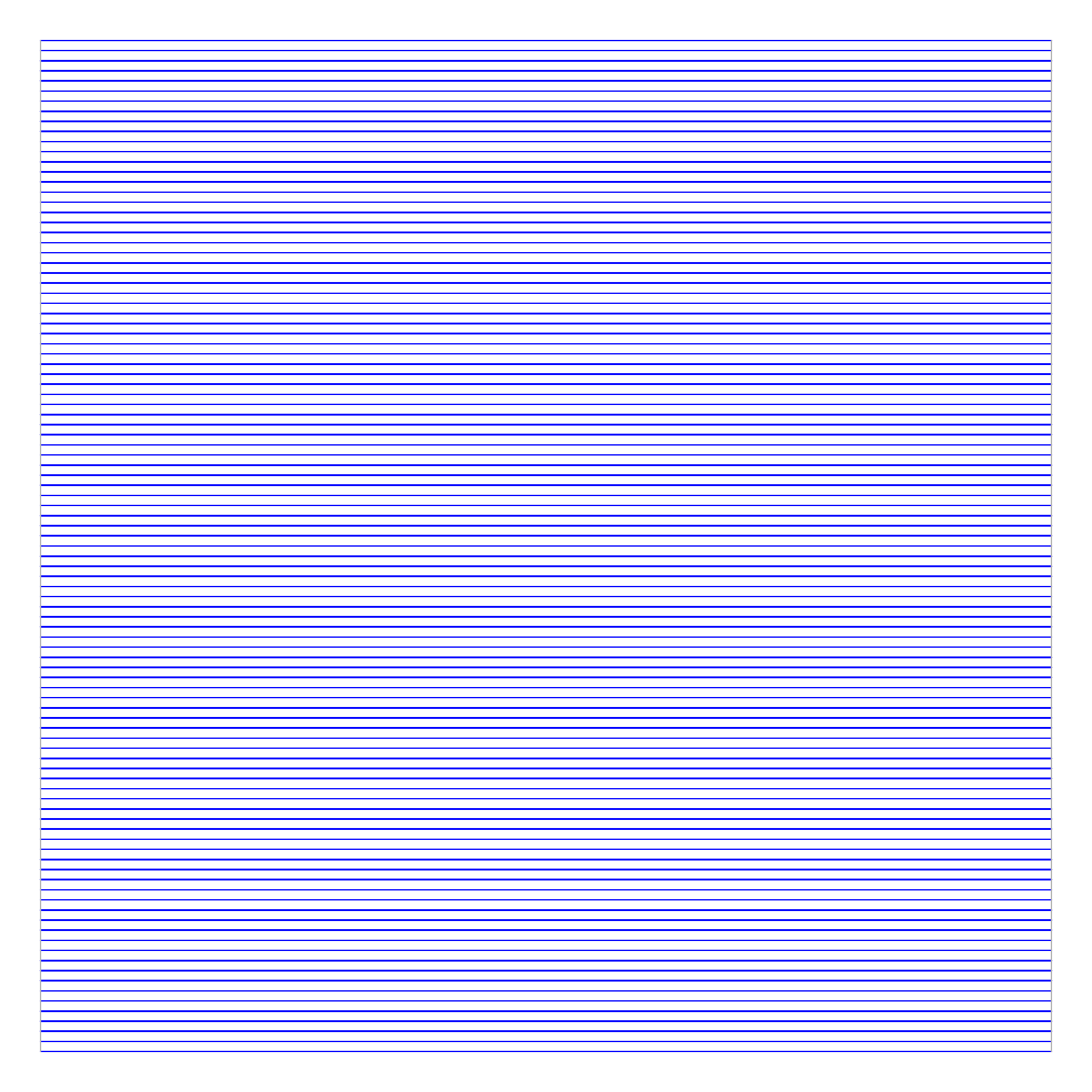}
\caption {$y=x$}
\label{fig:par:posline}
\end{subfigure} 
\begin{subfigure}{0.24\textwidth}
\centering
\includegraphics[width=\textwidth, angle=0]{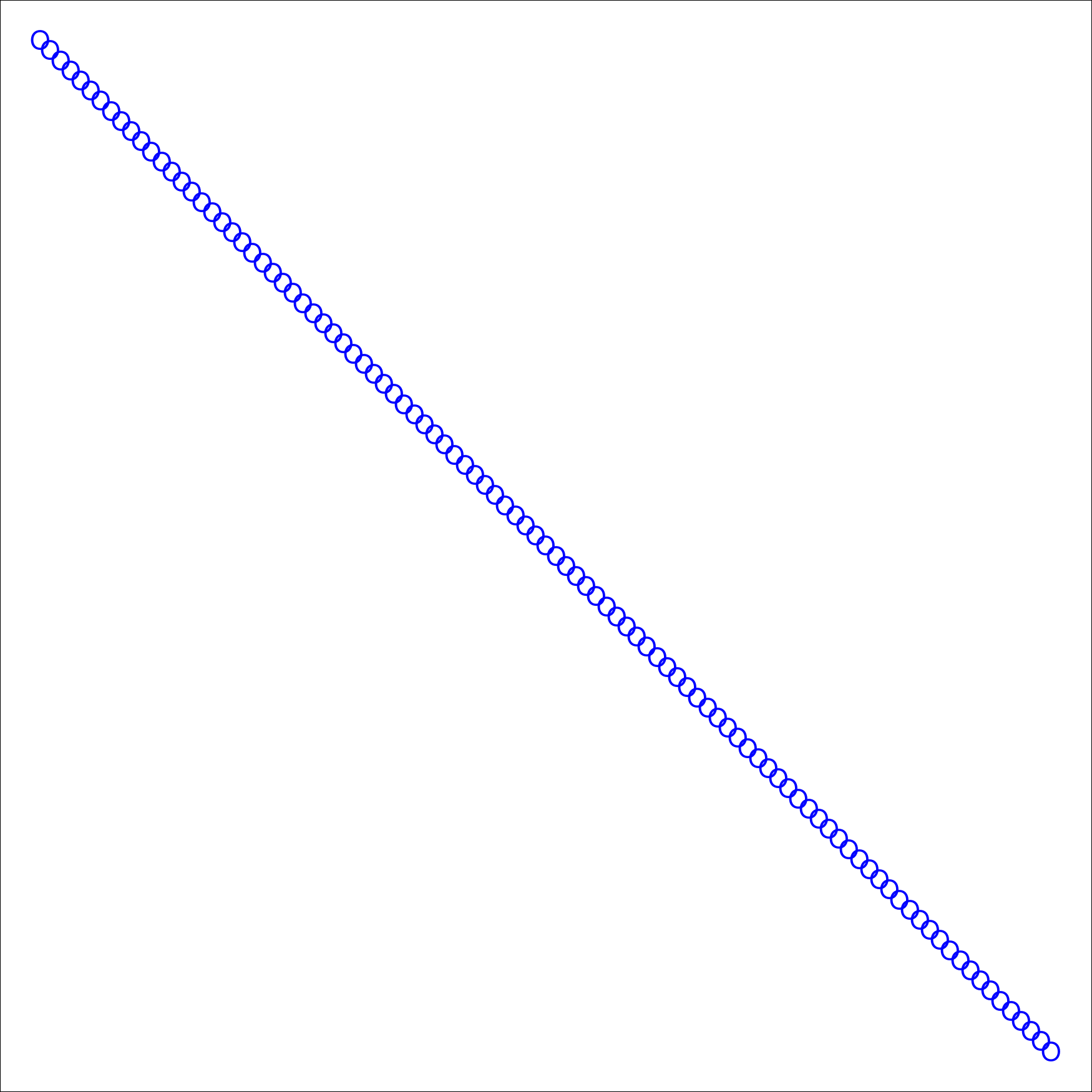}\\
\includegraphics[width=\textwidth, angle=0]{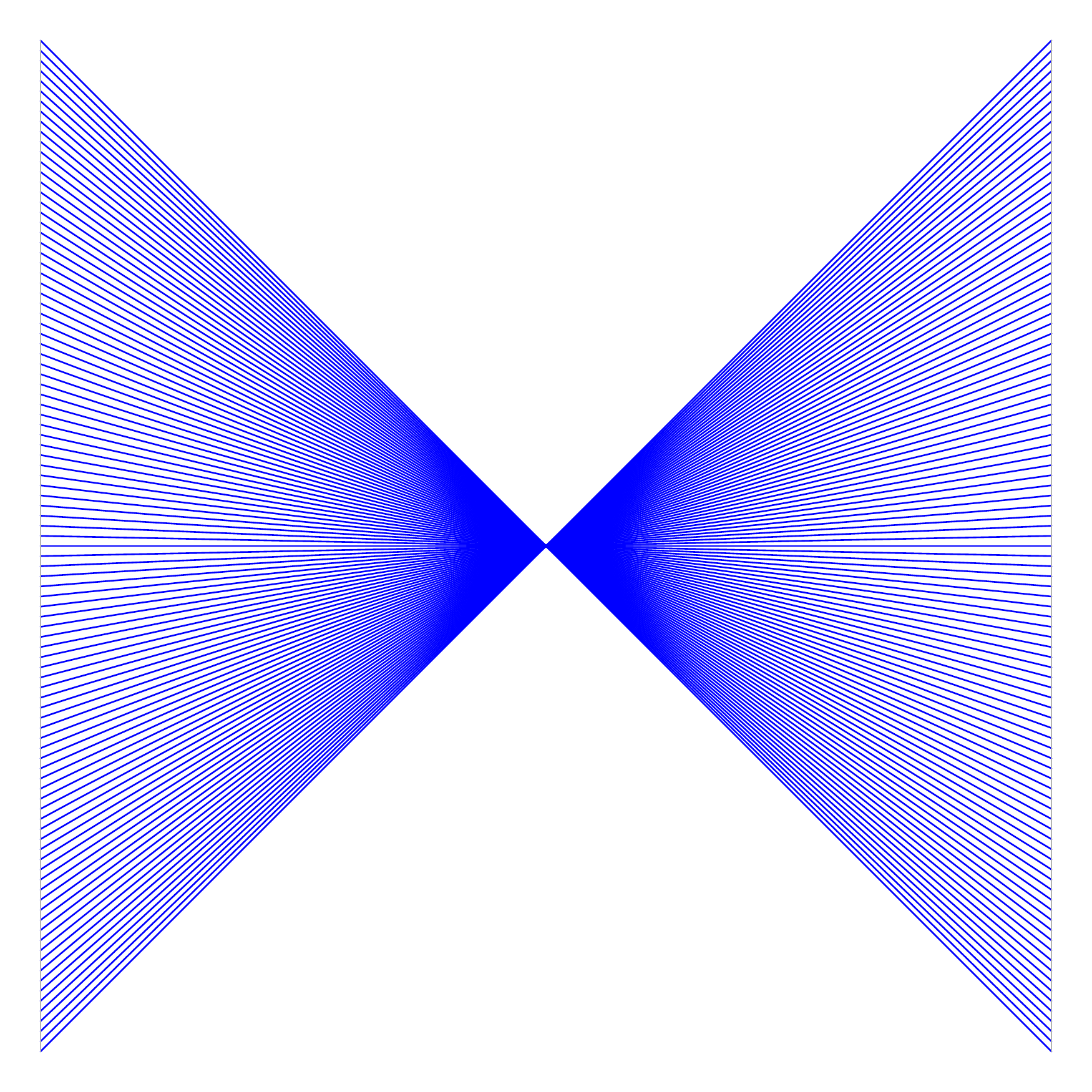}
\caption {$y=-x$}
\label{fig:par:negline}
\end{subfigure} 
\begin{subfigure}{0.24\textwidth}
\centering
\includegraphics[width=\textwidth, angle=0]{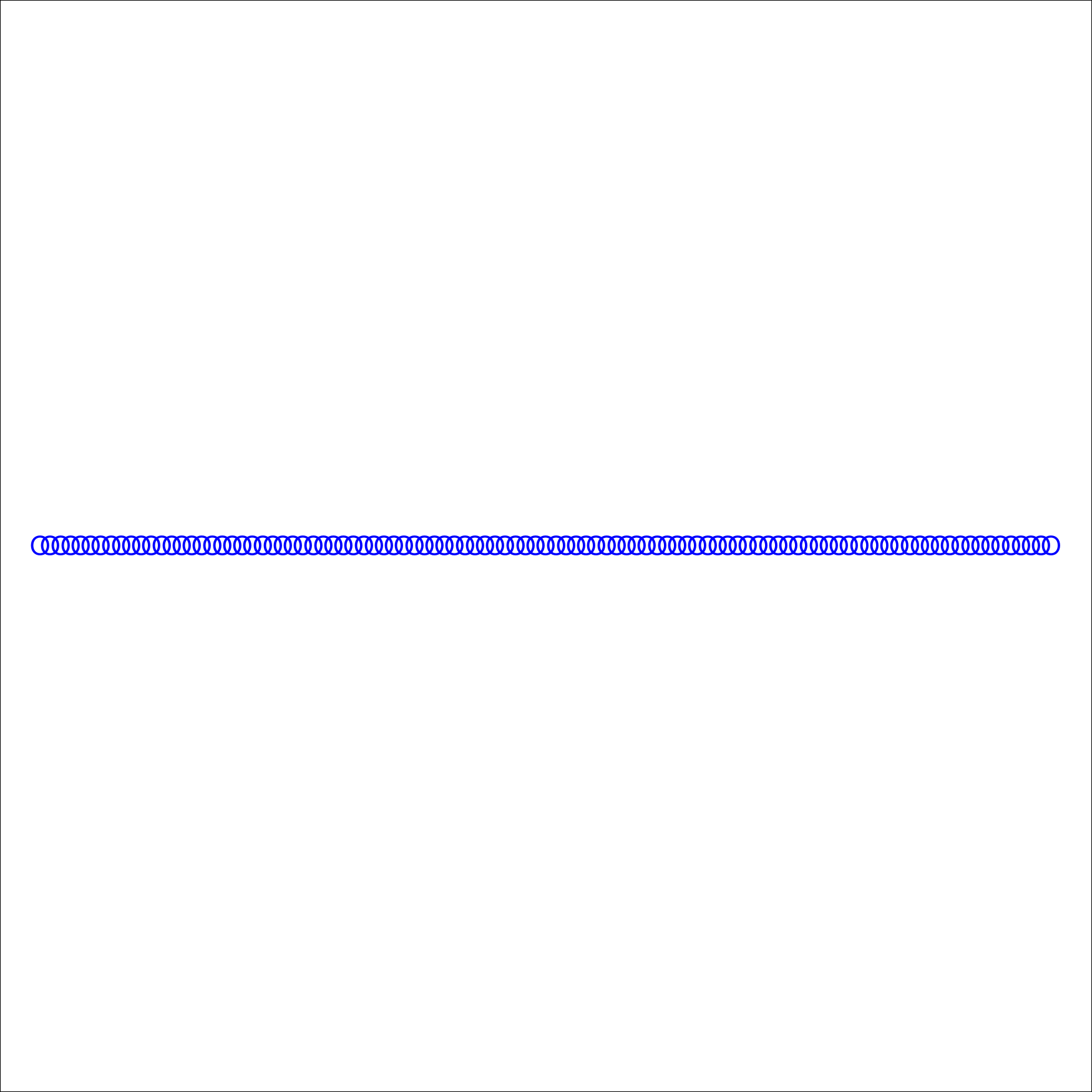}\\
\includegraphics[width=\textwidth, angle=0]{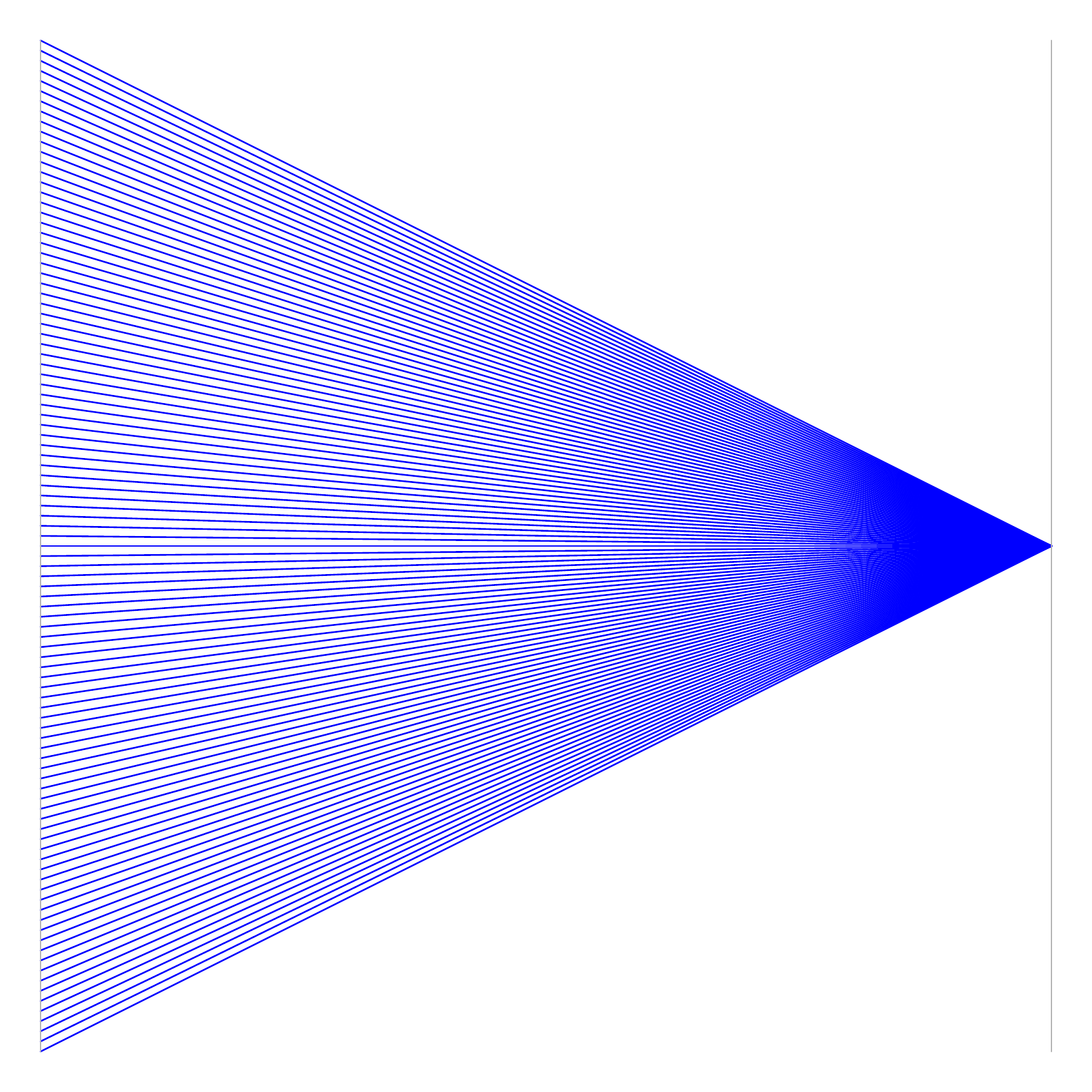}
\caption {$y=0$}
\label{fig:par:cste}
\end{subfigure} 
\begin{subfigure}{0.24\textwidth}
\centering
\includegraphics[width=\textwidth, angle=0]{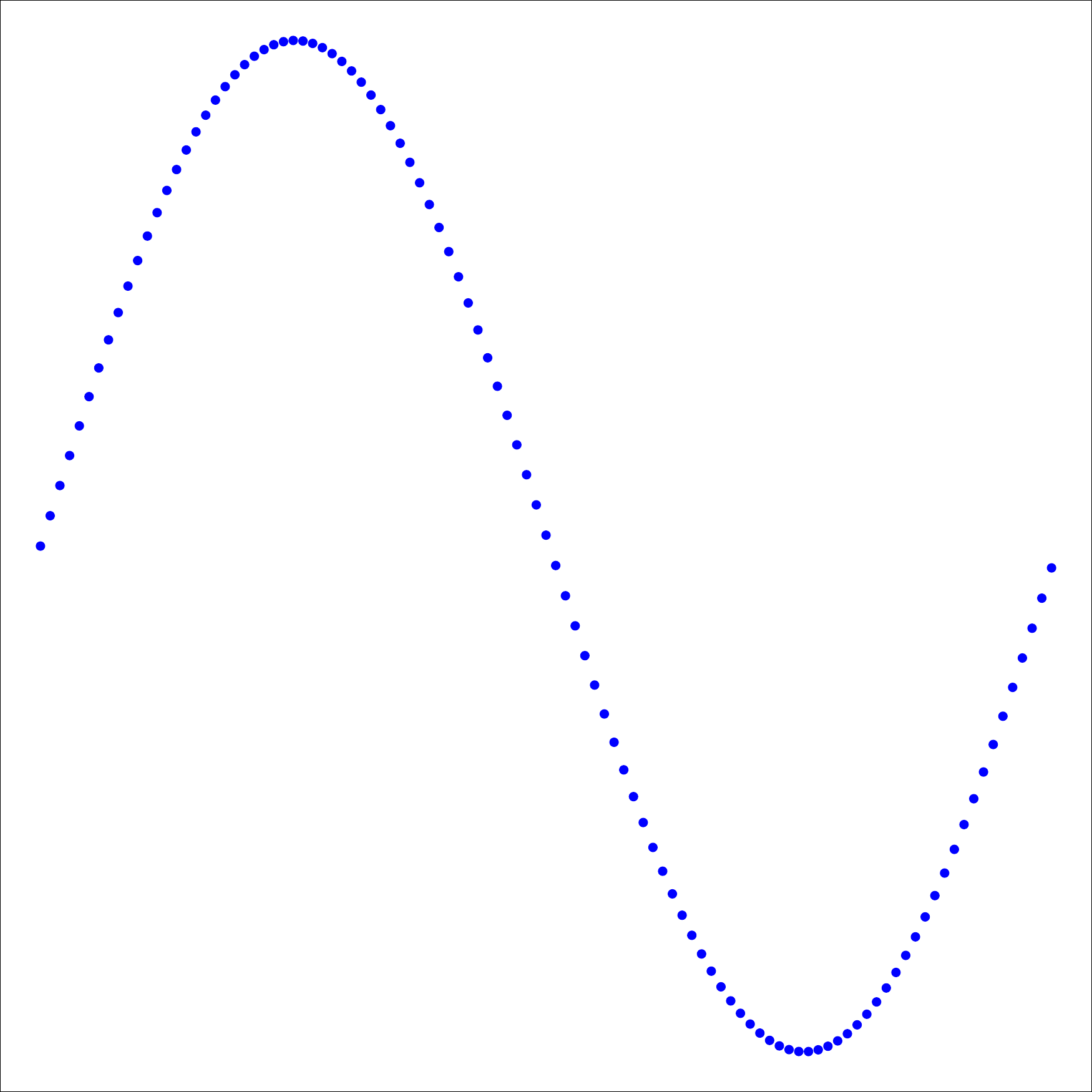}\\
\includegraphics[width=\textwidth, angle=0]{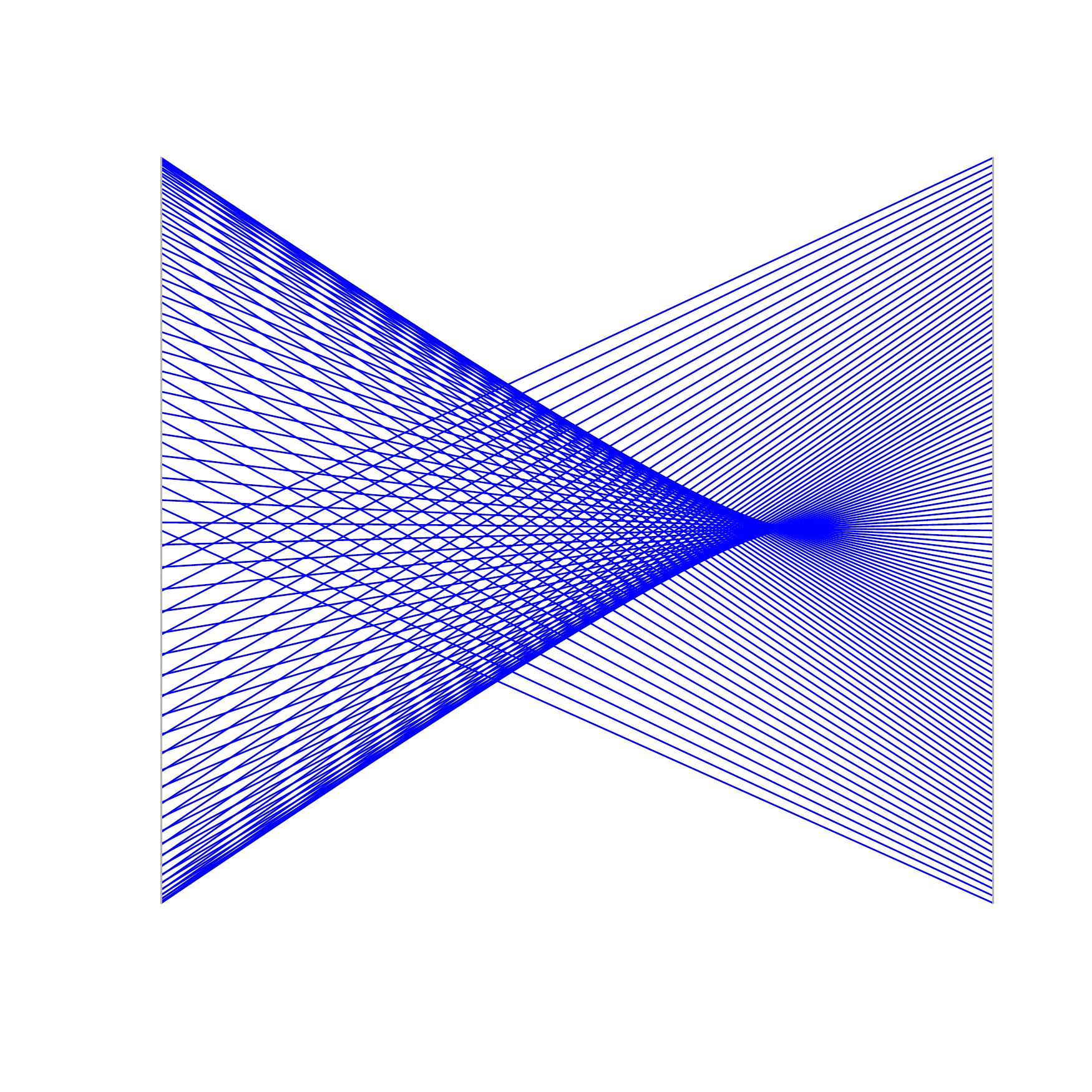}
\caption {$y=\sin(x)$}
\label{fig:par:sin}
\end{subfigure}

\caption{The duality between the Orthogonal coordinates (top) and the parallel coordinates (bottom) for $4$ common functions.}
\label{fig:lin_duality}
\end {figure*}

Coordinate reordering helps highlighting data dependencies, promotes visual data mining, and facilitates data exploration. Figure~\ref{fig:clutterorder} shows an example of four-dimensional data in its original order and after being reordered properly. In Figure~\ref{fig:ordercorrelations} there is a linear relation between $x_1$ and $x_3$ which is not visible in Figure~\ref{fig:withoutorder}. This relation is detected through many parallel lines between the two coordinates. Further examples on other dependence are presented in Figure~\ref{fig:lin_duality}. Interactive software enable manual attribute reordering. Users can change the order of attributes by switching axes. Handling the order manually is time consuming, but still some important relationships  may remain undetected. Developing an automatic technique seems essential for a good visualization, specially for large number of attributes.

Some authors proposed automatic techniques to find the best order for data visualization. The proposed techniques focus on highlighting the dependence among attributes. They aim to put an  attribute in the neighborhood of the most dependent attribute. For instance, \citet{ankerst1998similarity} proposes a technique to minimize the dissimilarities or partial dissimilarities between two adjacent attributes. The dissimilarity is often measured through the Euclidean distance over a pair of standardized attributes. It is not difficult to see that minimizing the Euclidean distance coincides with maximizing the squared correlation.
Unfortunately, correlation (or Euclidean distance) is unreliable to uncover all types of dependencies. Correlation is a deficient measure to uncover nonlinear dependencies \citet{cellucci2005MI}. For instance if $x$ follows a symmetric distribution such as Gaussian, the correlation between $x$ and $x^2$ is zero. \cite{Lu:SVD:NCC} suggests  some re-ordering techniques based on nonlinear correlation.

\citet{peng2004cluttercluster} proposes another technique for coordinate reordering, which aims to reorder by minimizing outliers between two neighboring coordinates. An observation is considered as an outlier if it involves no neighboring data. The neighbor is defined by the Euclidean distance after applying a certain threshold. This technique is sensitive to the chosen threshold. \citet{Johansson2009reordering} suggest a reordering technique using variety of metrics, e.g. maximizing correlation, reducing the number of clutters, etc. This approach gives an effective visualization and exploration of structures within a large multivariate data set, and meanwhile provides enhancement of diverse structures by supplying a range of automatic variable orderings.
\citet{Ferdosi2011reordering} suggest subspace clustering and coordinate ranking. \citet{Dasgupta2010pargnostics} propose several reordering metrics such as number of crossing lines, angle of crossing, and mutual information.
Another algorithm is independently developed using a genetic algorithm, to highlight important features and  allow the detection of irregularities using Pearson correlation in  \citet{boogaerts:order:2012:gene}. \citet{Lu:SVD:NCC} combine singular value decomposition to select the attributes that have the highest contribution and then applies a nonlinear correlation coefficient to order the axes.
Many techniques are proposed to improve the visualization of data in parallel coordinates. Each technique suggests an order based on a specific criterion. Our framework is general and is able to contain different measures.


The perception of patterns and clusters depends on the choice of the coordinate system. Therefore, it is important to know how to read the coordinate system. Despite the spread of parallel coordinates between practitioners, it is still unknown to many researchers in academica, especially when it comes to the interpretation of the shapes observed in parallel coordinates. Some authors has shown interests in studying the transformation from the orthogonal coordinates to the parallel coordinates. \citet{Inselberg1985} states that the representation in parallel coordinates is a projective transformation of orthogonal coordinates. \citet{heinrich2013state} study the transformation of a linear function to parallel coordinates in more details.

Some other dualities are studied by \citet{Inselberg1985} and \citet{Wegman1990ParShapes}. In point-line duality,  some other mappings can be expressed using the envelope of lines in parallel coordinates \citep{heinrich2013state}. Here we do not review the mathematical details, but rather focus on visual aspects. In Figure~\ref{fig:lin_duality}, some common functions are drawn in orthogonal coordinates and in parallel coordinates.

A set of points located on a line is represented in parallel coordinates by a set of lines that intersect at a definite point. The horizontal position of this point depends on the slope of the linear function. If the slope is negative, the intersection point is located between the parallel axes (Figure~\ref{fig:par:negline}). Different patterns are observed in a linear function with a slope superior to $1$, or inferior to $1$. However, as most software normalize the data, only parallel lines appears for a positive slope. This is illustrated in Figure~\ref{fig:par:posline} and Figure~\ref{fig:par:negline}.

Figure~\ref{fig:par:cste} shows a constant function. This function is illustrated by a set of lines that converge to a single point. A periodic function is translated by two sets of lines, intersecting in two different points as in Figure~\ref{fig:par:sin}.
Detecting the functions using the parallel coordinate shapes is still confusing, because some shapes resemble. The main difference is in the intersecting points. Despite the vagueness  in the interpretation of shapes, it is clear that when two attributes are dependent, the parallel coordinate graph shows a certain pattern.

\begin{figure*}[t]
\centering
\begin{subfigure}{0.2\textwidth}
\centering
\includegraphics[width=\textwidth, angle=0]{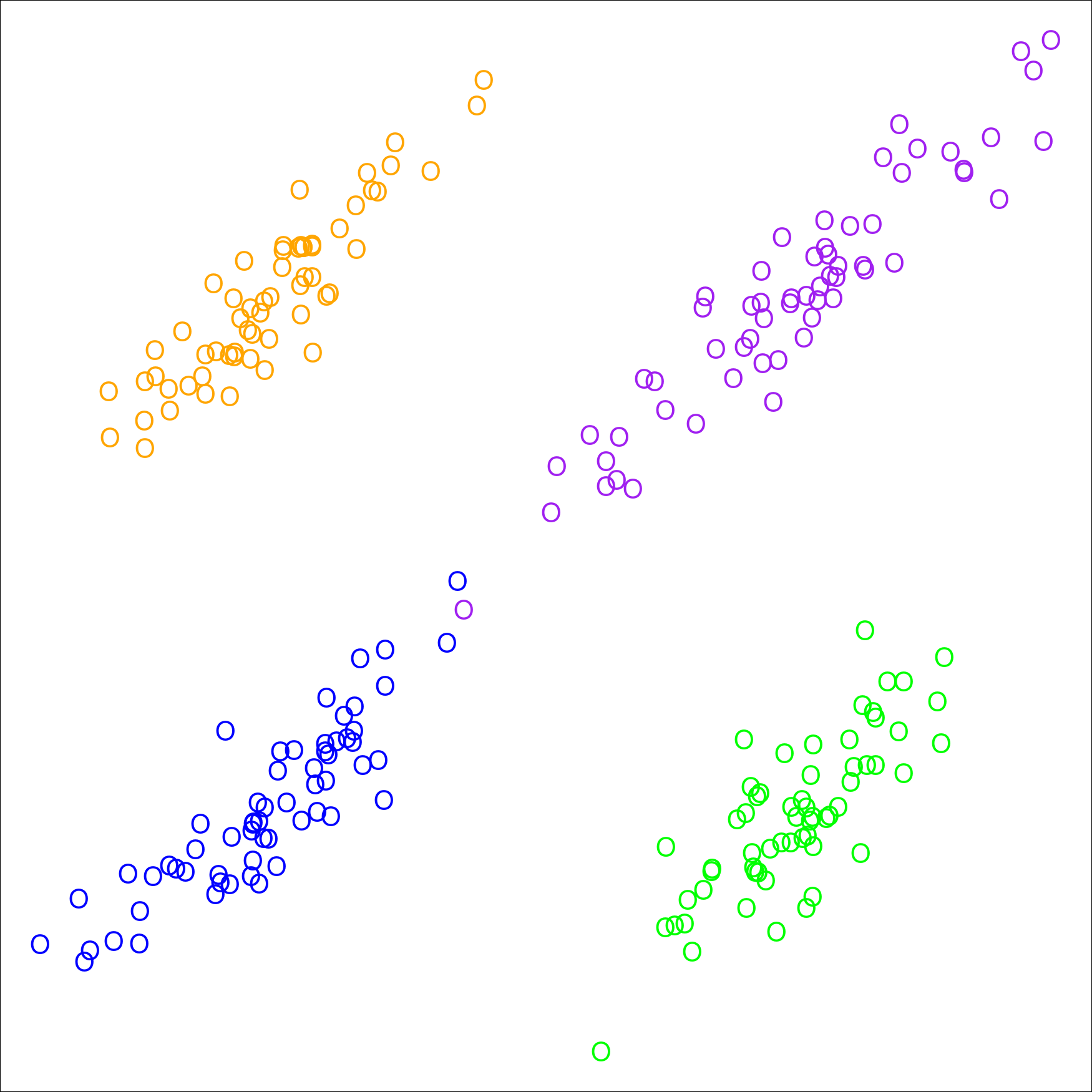}\\
\includegraphics[width=\textwidth, angle=0]{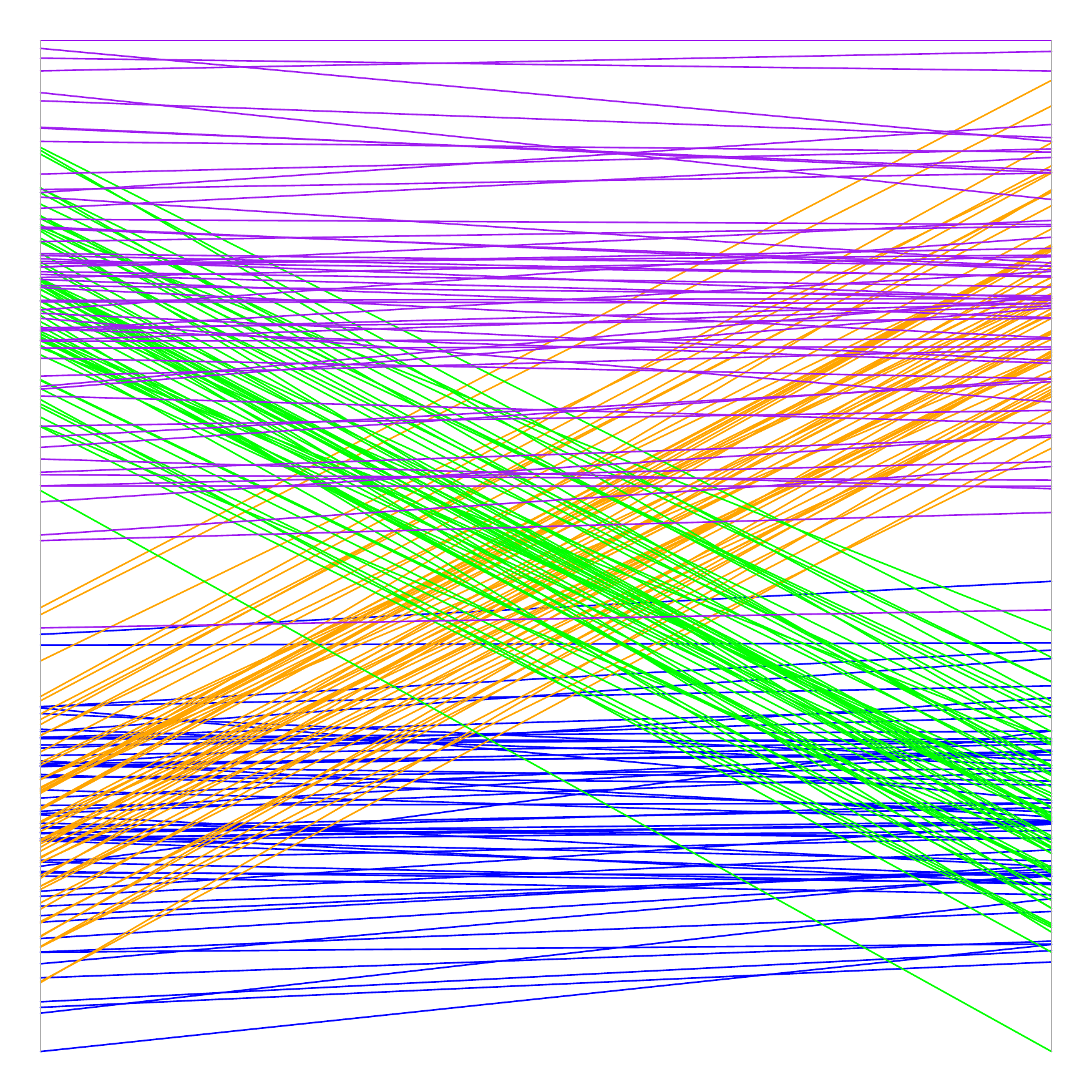}
\caption {Separable and correlated data}
\label{fig:par:sep:cor}

\end{subfigure} 
\begin{subfigure}{0.2\textwidth}
\centering
\includegraphics[width=\textwidth, angle=0]{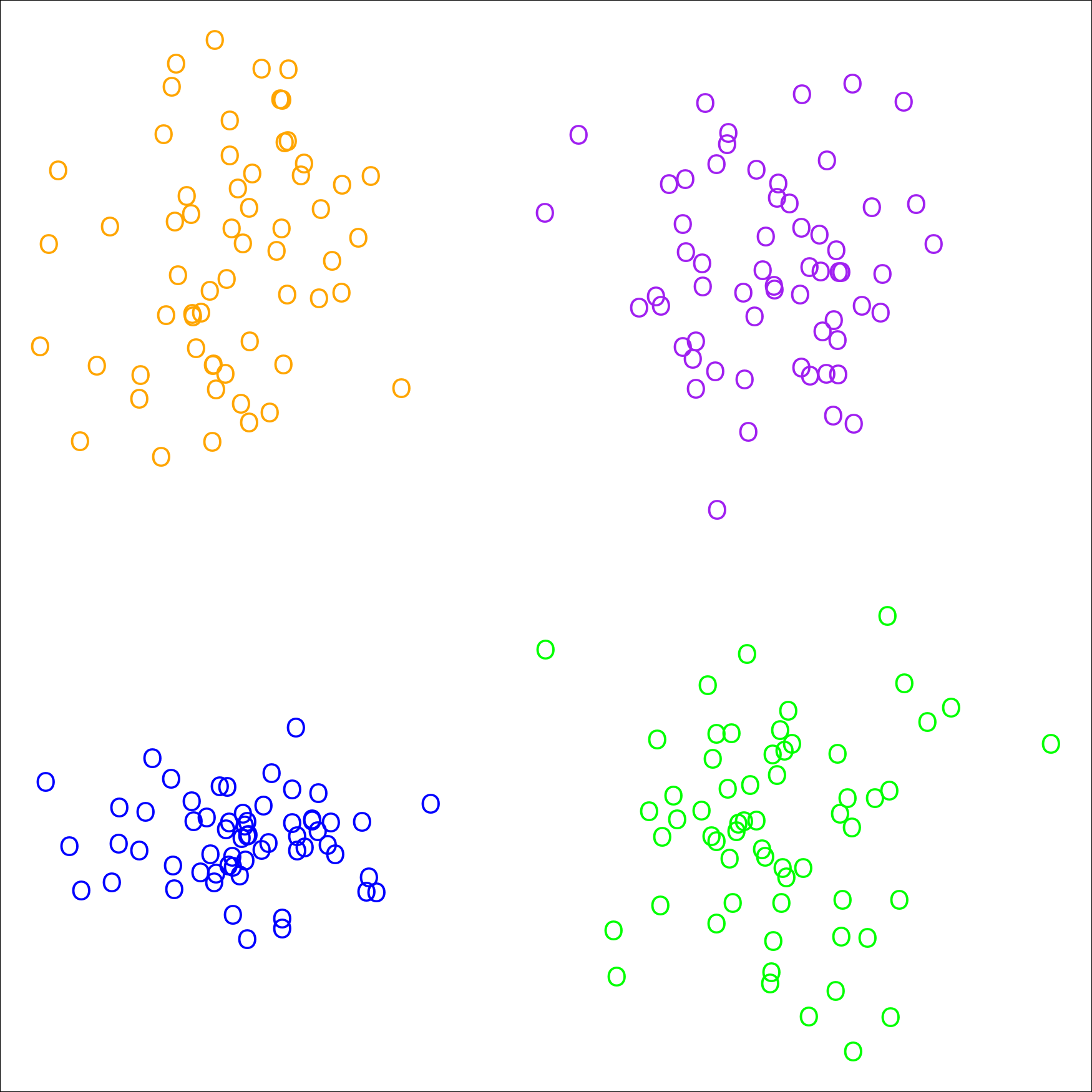}\\
\includegraphics[width=\textwidth, angle=0]{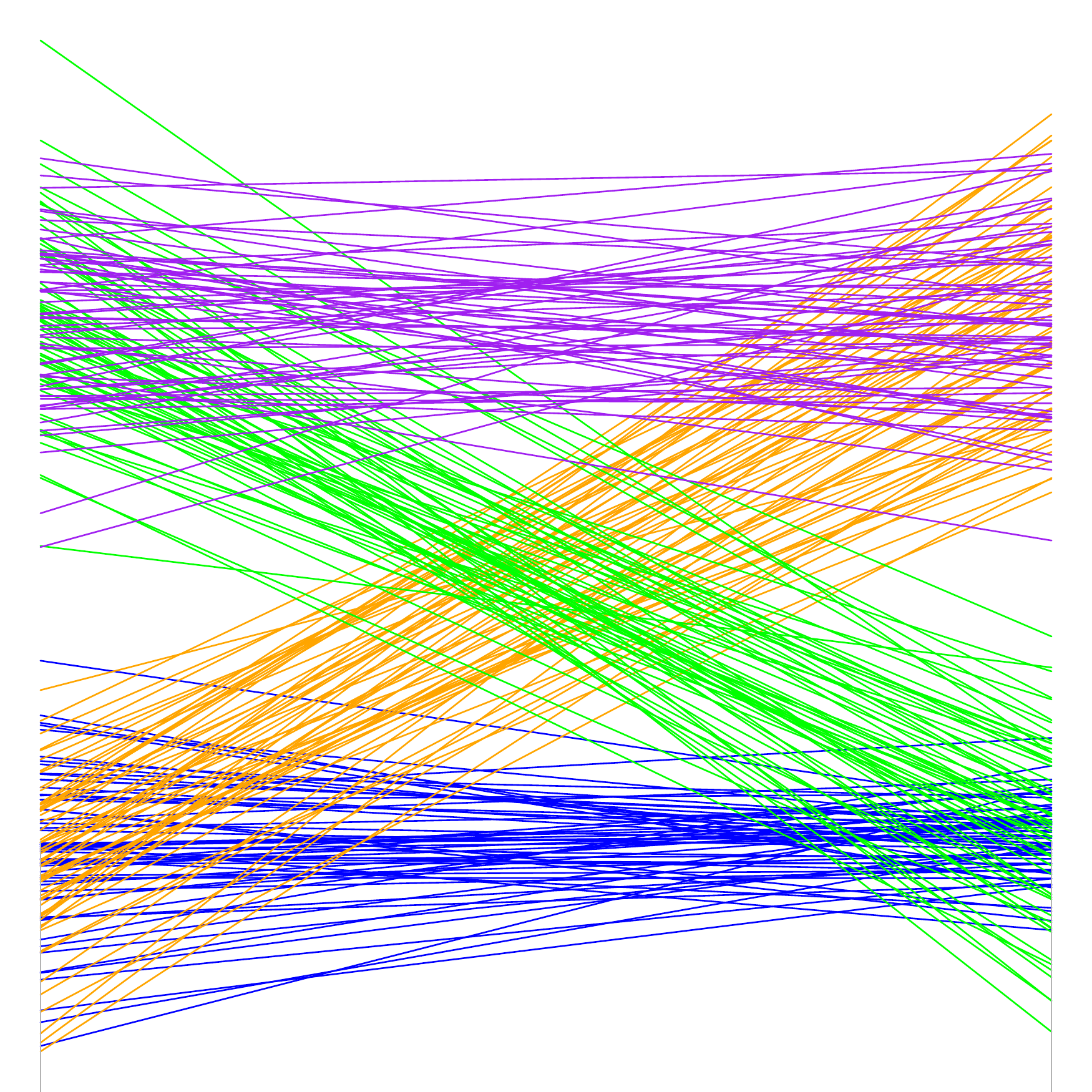}
\caption {Separable non correlated data}
\label{fig:par:sep:noncor}
\end{subfigure} 
\begin{subfigure}{0.2\textwidth}
\centering
\includegraphics[width=\textwidth, angle=0]{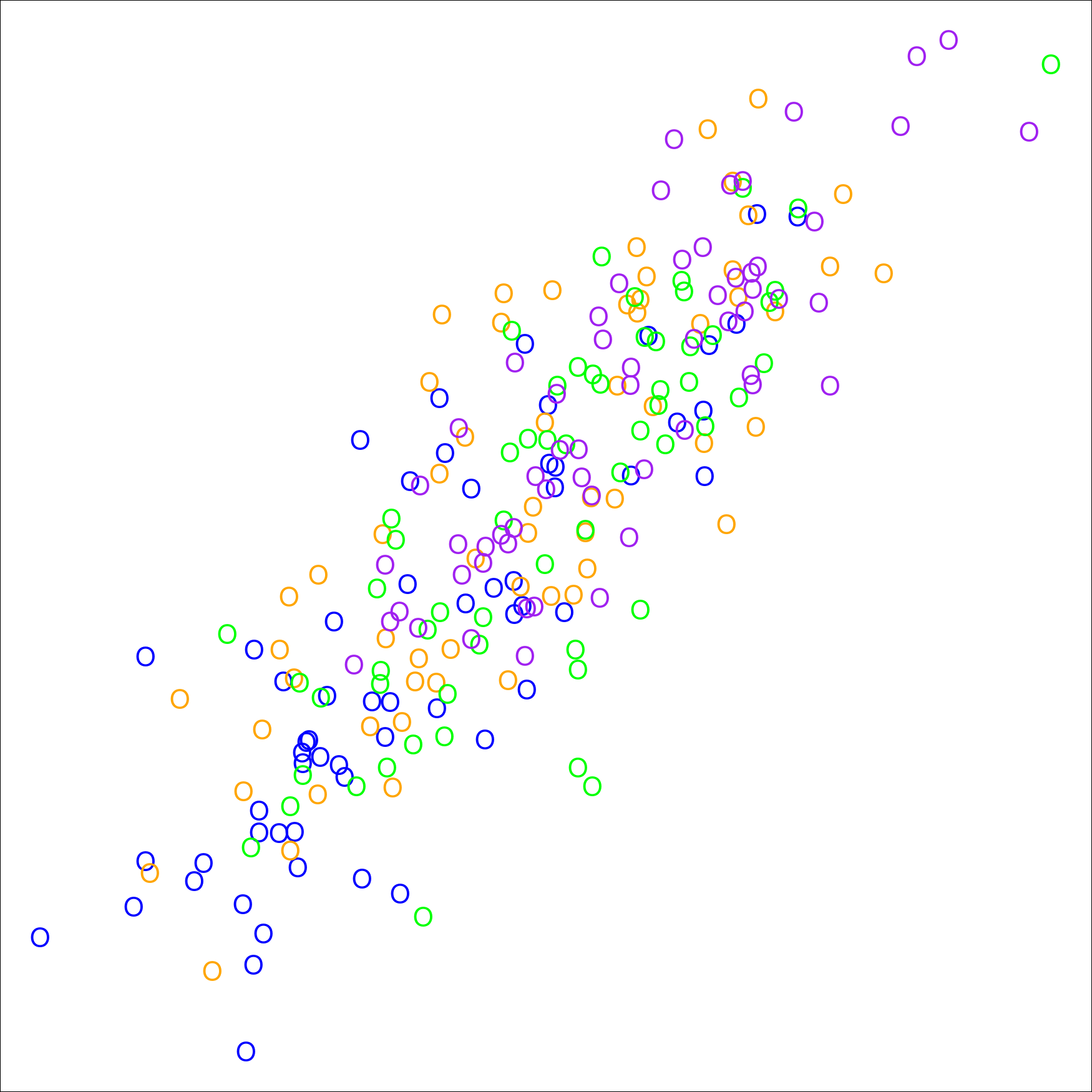}\\
\includegraphics[width=\textwidth, angle=0]{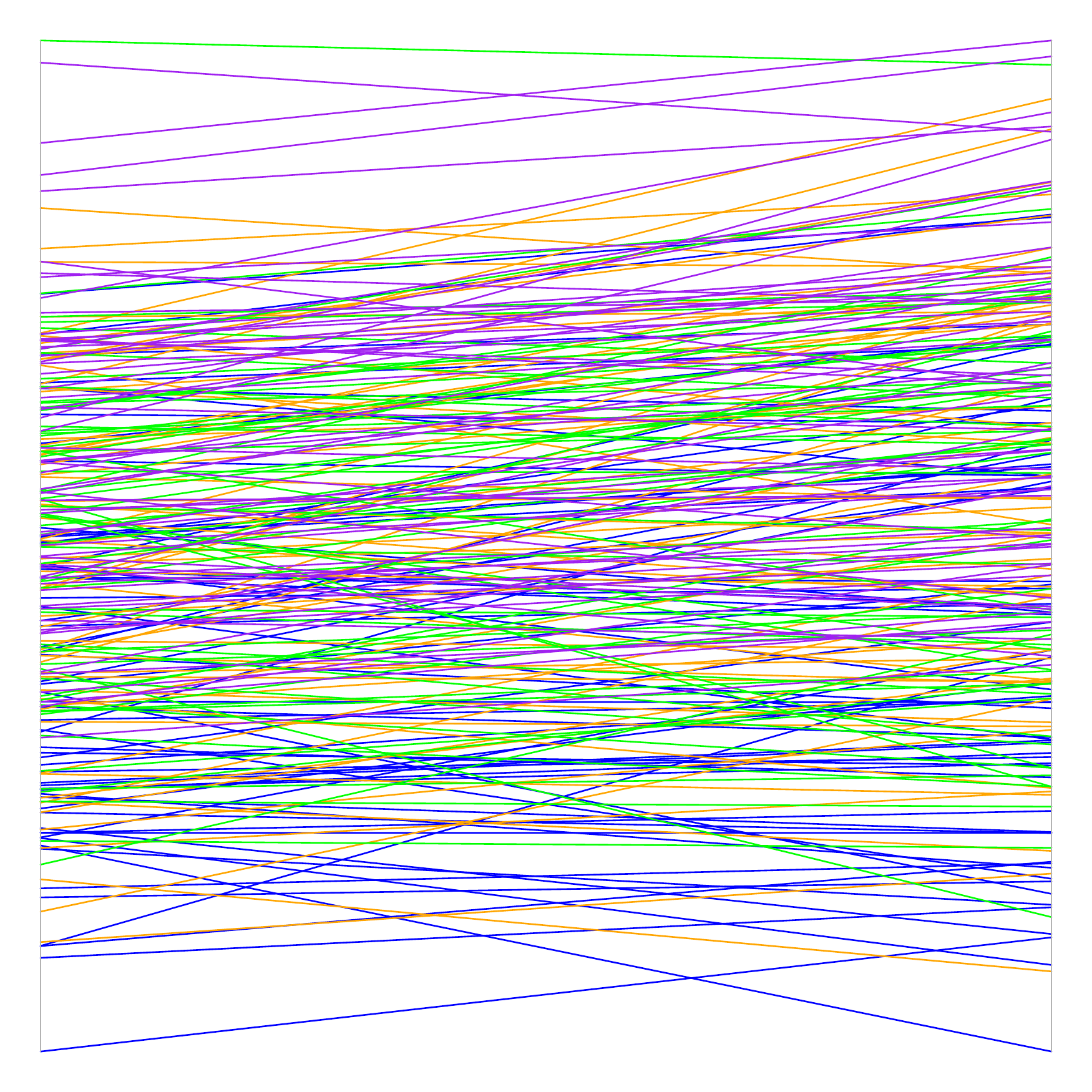}
\caption {Non separable correlated data}
\label{fig:par:nonsep:cor}
\end{subfigure} 
\begin{subfigure}{0.2\textwidth}
\centering
\includegraphics[width=\textwidth, angle=0]{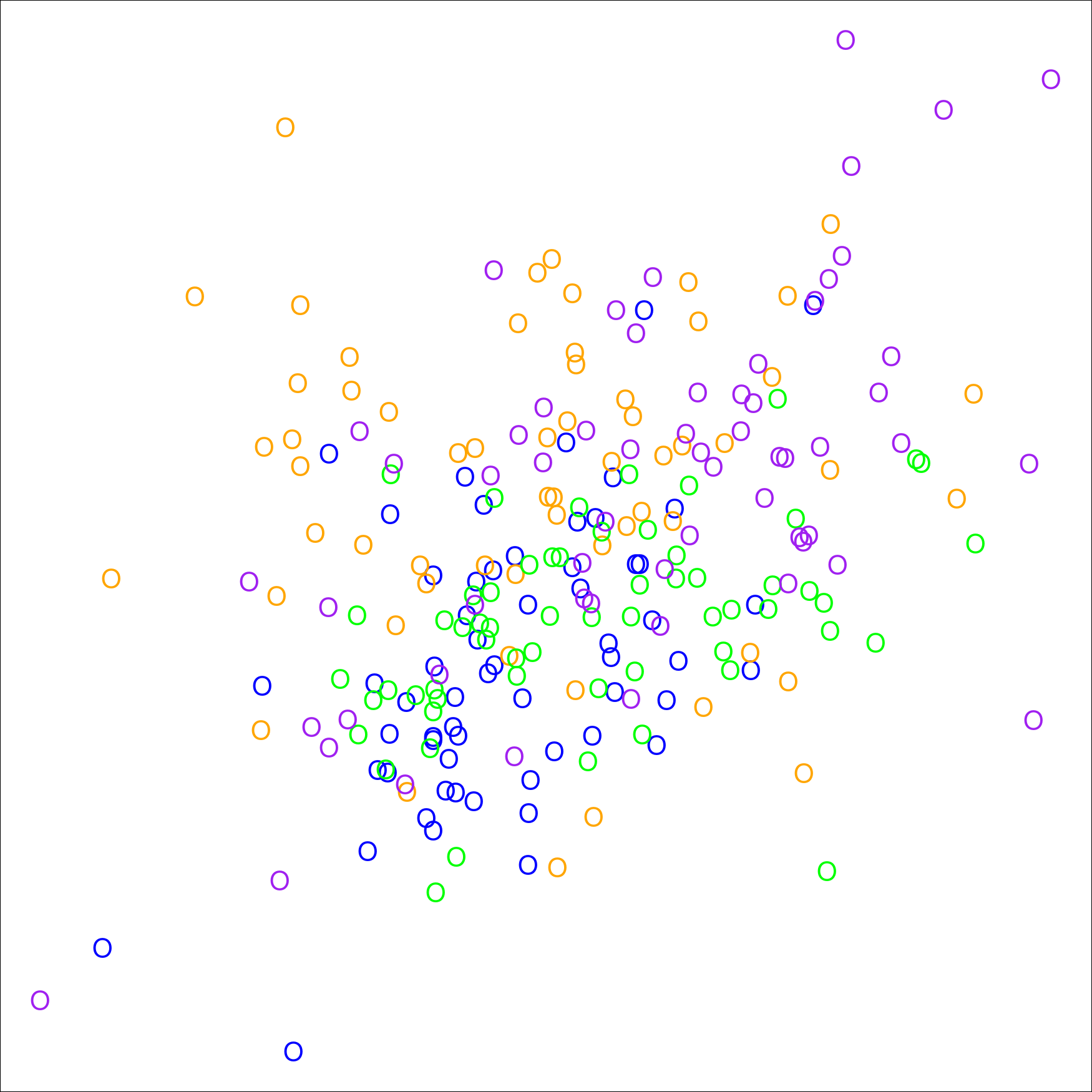}\\
\includegraphics[width=\textwidth, angle=0]{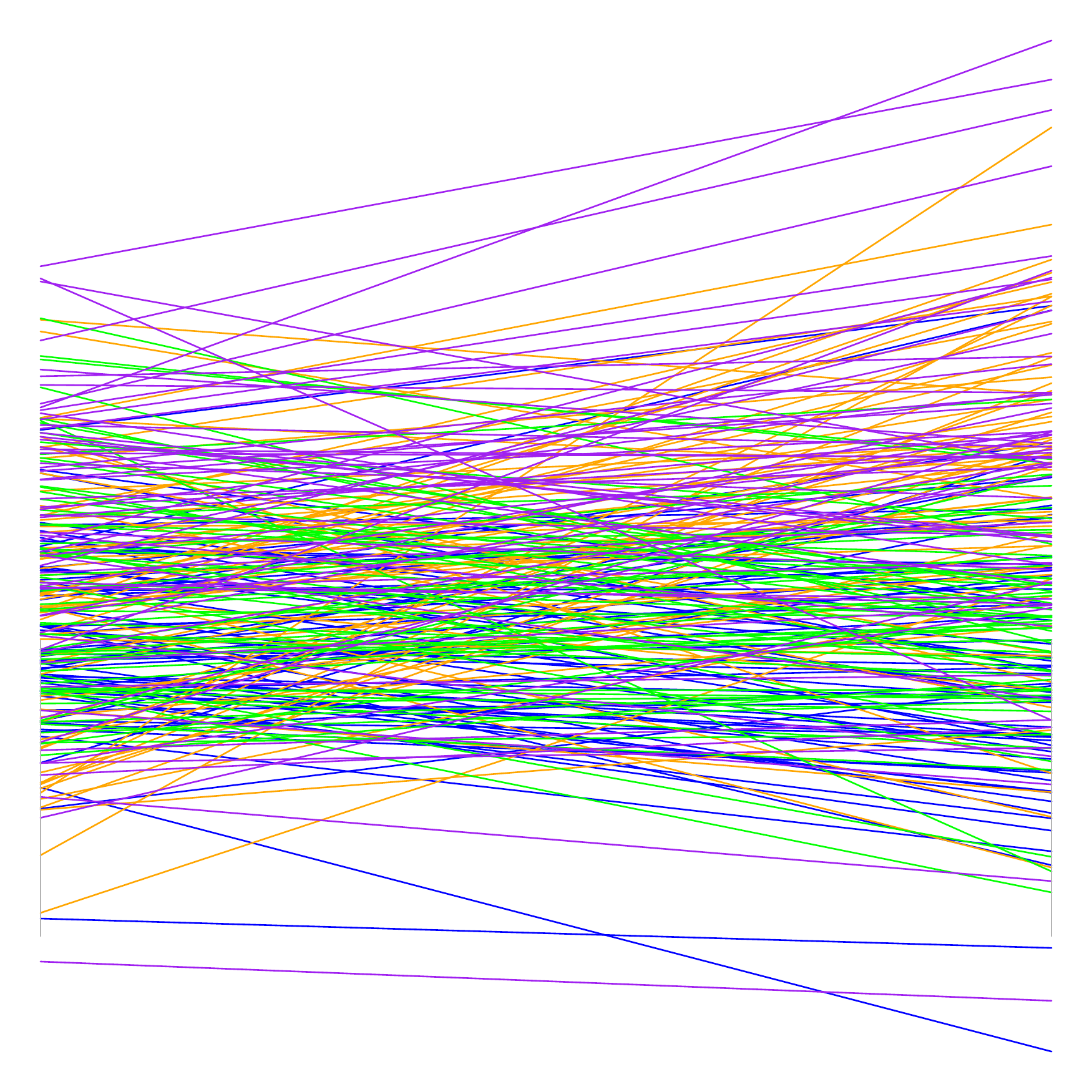}
\caption {Non separable non correlated data}
\label{fig:par:nonsep:noncor}
\end{subfigure}

\caption{Separation and correlation in orthogonal coordinates (top panel) and in parallel coordinates (bottom panel).}
\label{fig:clusters_duality}
\end {figure*}

Cluster visualization is different in orthogonal coordinates compared with parallel coordinates. Figure~\ref{fig:clusters_duality} illustrates the separation and correlation in both coordinate systems. Figure~\ref{fig:par:sep:cor} shows separable and correlated data. The clusters are visible and some patterns appears in parallel coordinates. These patterns translate a set of linear functions with different coefficients to a set of lines. Figure~\ref{fig:par:sep:noncor} shows separable and uncorrelated data. The patterns are not much different than Figure~\ref{fig:par:sep:cor}, but the clusters become more distinguishable. Figure~\ref{fig:par:sep:noncor} translates correlated but on non-separable data, and Figure~\ref{fig:par:nonsep:noncor} illustrates uncorrelated  and non-separable data.
Despite the difference in cluster perception in orthogonal coordinates and parallel coordinates, when data are separable, the clusters are distinguishable in both coordinate systems.

\section{Dimension Reordering}
In this section, we present a general framework for dimension reordering. Subsection~\ref{sec:criterion} defines the general bivariate measure (general information) and Subsection~\ref{sec:optim} explains the optimization algorithm proposed to find an optimal order.
\subsection{General Information Criterion}
\label{sec:criterion}
Various methods are used to order coordinates, from Euclidean distance to correlation. As only two coordinates are visualized at a time, it looks promising to order coordinates through some measures defined over the bivariate data distribution. Take two arbitrary attributes, say $x_1, x_2$. Define two hypothetical bivariate probability measures over the product of their sample space, and over the same sigma algebra $\mathcal F$. In other words, define two distinct but comparable probability spaces $\big(\Omega, \mathcal{F},F \big),$ and $\big(\Omega, \mathcal{F},H \big)$ for $(x_1,x_2)$. For the simplicity of notation we denote the probability measures $F$ and $H$ by their imposed distribution functions $F(x_1,x_2)$ and $H(x_1,x_2)$. Let 
$F(x_1, x_2)$ and $H(x_1,x_2)$ impose different probability measures, i.e. $$\exists (x_1, x_2) \in \real^2 \textrm{ such that }F(x_1, x_2)\neq H(x_1, x_2).$$ 
Define the \emph{general information} as
\begin{equation}
\GI(x_1, x_2)= {1 \over {G''(1)}} \int \int G\left\{ { dF(x_1, x_2)\over dH(x_1, x_2) }\right\}dH(x_1,x_2)\label{eq:GI},
\end{equation}
where $dF(x_1, x_2)/dH(x_1, x_2)$ is the Radon-Nikodym derivative, $G(.)$ is a univariate smooth function and $G''(1)\neq 0$ is the second derivative of $G(.)$ at $1$. The second derivative  in \eqref{eq:GI} adjusts for scaling. The criterion defined in \eqref{eq:GI} is closely related to the Kullback-Leibler and phi divergence, the cross entropy, and the joint entropy.

The choice of $F$ relative to $H$ defines the measuring concept and the choice of $G(.)$ defines the measuring statistic. A common choice of $F$ and $H$ is the data joint distribution and the product of marginal data distributions, respectively. In this case, the measuring concept reduces to dependence.  The Pearson correlation as a measure of dependence arises if $F(x_1,x_2)$ is bivariate Gaussian.

A common choice of $G(.)$ is $G(u)= u \log(u)$ which brings the Kullback-Leibler divergence of $F$ relative to $H$.  Our suggestion for $G(u)$ is a univariate function that 
\begin{enumerate}[label=\roman*)]
\item vanishes at 1, i.e. $G(1)=0$,
\item its first derivative is smooth at 1 , i.e. $G''(u)$ is bounded in an infinitesimal neighborhood $u \in (1-\epsilon,1+\epsilon)$.
\end{enumerate}
The first condition ensures that $\GI$ is well-defined. In other words, $\GI = 0$ if and only if the reference probability measures $F$ and $H$ coincide. The second condition ensures the asymptotic statistical behavior as the number of observations $n\to\infty$, see Theorem~\ref{theo:asymp}.

One may choose the statistic of interest by varying $G(u)$. It is easier to understand the role of $G(.)$ in the context of discrete random variables. If $(x_1, x_2)$ is a pair discrete random variables, $H(x_1, x_2)=F(x_1)F(x_2)$, then various famous statistics of contingency tables are derived by varying $G(u)$  
\begin{itemize}
\item $G(u)=2 u\log u$ gives the log likelihood-ratio statistic,
\item $G(u)=(u-1)^2$ gives the Pearson chi-square statistic,
\item $G(u)=u(1-1/\sqrt{u})$ gives the Freeman-Tukey statistic,
\item $G(u)=(1-u)^2/u$  gives the Neyman statistic,
\item $G(u)=u(\sqrt[3]{u^2}-1)$ gives the Cressie-Read statistic,
\end{itemize} 
and more importantly  $G(u)=u\log u$ is the mutual information
$$\GI(x_1,x_2)=\sum_{x_1}\sum_{x_2} p(x_1,x_2) \log {p(x_1,x_2)\over p(x_1)p(x_2)},$$ 
where $p(x_1,x_2)$ is the joint probability mass, $p(x_1)$ and $p(x_2)$ are the marginal masses.  Under some mild assumptions all of the above statistics follow a scaled chi-square distribution.

Now we explore the asymptotic behavior of $\GI$. Define $$\hat\GI_n= {1 \over {G''(1)}} \int \int G\left\{ { dF_n(x_1, x_2)\over dH_n(x_1, x_2) }\right\}dH_n(x_1,x_2),$$ where $F_n$ and $H_n$ are the empirical distribution functions and
\begin{theorem}
\label{theo:asymp}
Suppose $F_n(x_1,x_2)$ and $H_n(x_1,x_2)$ are the empirical distribution functions that uniformely converge to $F(x_1, x_2)$ and $H(x_1,x_2)$. Assume $F(x_1, x_2)=H(x_1, x_2)$ almost surely, and 
\begin{itemize}
\item $\forall n,$ $F_n$ is measurable  with respect to the Lebesgue-Stieltjes measure $H_n$ .
\item If $dH_n=0$, define ${dF_n \over dH_n}=1$.
\item $G'''(u)$ is uniformly bounded on $u\in(1-\epsilon, 1+\epsilon)$
\item $\forall n, H_n$ is nested in $F_n,$ i.e. $\nu = \mathrm{dim}(F_n) - \mathrm{dim}(H_n)>0,$
\end{itemize}
then 
$$2n{\GI}_n(x_1, x_2) \leadsto \chi^2_\nu \mathrm{~~as~~} n\to\infty.$$
\end{theorem}

Measures that coincide with contingency tables quantify dependence. It is more meaningful to measure the separation instead of dependence if visualization towards data clutter is the target. Therefore, one may define $F(x_1, x_2)$ to be a $k$ component distribution 
\begin{equation}
dF(x_1,x_2)=\sum_{c=1}^k p_c g_{\boldsymbol\mu_c}(x_1,x_2) dx_1 dx_2
\label{eq:kcomp}
\end{equation}
and $H(x_1,x_2)$ to be a single component distribution
\begin{equation}
dH(x_1,x_2)= g_{\boldsymbol\mu}(x_1,x_2)dx_1 dx_2,
\label{eq:onecomp}
\end{equation}
where $g_{\boldsymbol\mu}(.,.)$ is a density family indexed by the location parameter $\boldsymbol\mu$.
Such a measure mimics the silhouettes \citep{Rousseeuw:Silhouette:1987} if $g$ is Gaussian bivariate density.

Ordering with respect to outliers is feasible through assigning a heavy-tailed, such as the Student's t-distribution, for $F$ and a bivariate Gaussian for $H$.  Many other concepts such as dispersion, non-linear correlation, skewness, prediction power, multi-collinearity, data classification  etc, can be quantified through the general information criterion \eqref{eq:GI}, and then be used to order the coordinates for further visual inspection.

\subsection{Order Optimization}
\label{sec:optim}
Suppose data contain $p$ attributes. The total number of coordinate permutations is $p!$ which is impossible to check visually for large $p$. It is natural to put the most informative coordinates early in the graph.  This is specially helpful while data are high-dimensional to visualize only the coordinates with maximum relevant information.

Suppose the general information matrix, call the symmetric weight matrix, is computed for all pairs of attributes $\mathbf W_{p\times p}=[w_{ij}]$, where $w_{ij}=\GI(x_i,x_j).$ The problem of finding optimal neighboring coordinates is reduced to estimation of a binary symmetric adjacency matrix $\mathbf A = [a_{ij}]$ that maximizes the total information
\begin {eqnarray} 
\hat{\mathbf A}&=& \argmax \lVert {\mathbf A \odot \mathbf W} \rVert\label{eq:objective} \\
\textrm{s.t.}&& \nonumber\\
a_{ij}=0 & \mathrm{or} & a_{ij}=1 \label{eq:const:binary}
 \\
 \mathbf a_{i}^\top \mathbf 1 &=& \mathbf 1 ^\top \mathbf a_{j} = 2 \label{eq:const:rowcol}\\
  a_{ij}&=& a_{ji}, \label{eq:const:symmetry}\\
\lVert \mathbf A \rVert &\leq & 2 q  \label{eq:const:select}
\end{eqnarray}
where $\mathbf a_{i}^\top$ is the $i$th row of $\mathbf A$, $\mathbf a_j$ is the $j$th column of $\mathbf A$, $\odot$ is the Hadamart  product, and $\lVert\mathbf A \rVert=\sum_i\sum_j \lvert a_{ij}\rvert $ is the \emph{$L_1$} Frobenius norm.

The objective function $\sum_{i=1}^p \sum_{j=1}^p a_{ij}w_{ij}$ in \eqref{eq:objective} computes the utility of incorporating some adjacent coordinates. The constraint \eqref{eq:const:binary} ensures whether or not a coordinate is neighbor to another. The constraint \eqref{eq:const:rowcol} ensures a coordinate is neighbor to only two other coordinates. The constraint \eqref{eq:const:symmetry} imposes symmetry on the adjacency matrix. The constraint \eqref{eq:const:select}, for a $q < p$, selects only $q$ out of $p$ coordinates for visualization.

Standard solvers such as CPLEX can be used to solve this integer-linear optimization program after fixing $q$. If $q\geq p$, the integer program only finds the adjacent coordinates and relaxes the selection. For high-dimensional data, this optimization is cumbersome to solve,  even with powerful computers. We propose a faster algorithm by optimizing the objective function \eqref{eq:objective} hierarchically as follows. 

The first pair of coordinates are the one that maximize the objective function at the first iteration 
\begin{eqnarray}
(\hat x_1, \hat x_2)&=& \argmax \GI(x_i,x_j) \label{eq:greedone}\\ 
1 \leq i \leq p-1& & i+1 \leq  j \leq p.  \nonumber
\end{eqnarray} 
The $j$th, $j=3,\ldots, q$ coordinates is 
\begin{eqnarray}
\hat x_j= \argmax  \GI(\hat x_{j-1},x_i),\label{eq:greedtwo} \\
i\in \{1,\ldots,p\}\backslash \{ \hat x_1, \hat x_2, \ldots,\hat x_{j-1} \}.\nonumber
\end{eqnarray}

The computation of this greedy algorithm is of time complexity $O(p^2)$ and dominated by the first step  of the algorithm \eqref{eq:greedone}. A faster algorithm of order $O(qp)$ can be achieved by fixing the first coordinate manually and ordering the remaining coordinates using \eqref{eq:greedtwo}. This technique is scalable with the number of coordinates $p$, specially for high-dimensional data while $q \ll p$.

\section{Application}
\label{sec:appl}
To test the proposed algorithm, we used two well-known datasets. The first is the white wine quality data  \citep{wine}. This dataset includes $12$ attributes. The second dataset is Golub genetic data \citep{Golub:1999un}. It is a high-dimensional data and only $q=50$ attributes out of $p=2030$ are selected for visualization.

\subsection{Wine Dataset}
\label{subsec:winedata}
These data are the result of a chemical analysis of white wines taken from \citet{wine}. The data include $n=4898$ measurements over $p=12$ attributes: fixed acidity ($x_1$), volatile acidity ($x_2$), citric acid ($x_3$), residual sugar ($x_4$), chlorides ($x_5$), free sulfur dioxide ($x_6$), total sulfur dioxide ($x_7$), density ($x_8$), pH ($x_9$), sulphates ($x_{10}$), alcohol ($x_{11}$) and quality ($x_{12}$, a score between $0$ and $10$).
This dataset is analyzed as a benchmark for outlier detection, classification, and regression. \citet{Dasgupta2010pargnostics} used this dataset to evaluate the dimension reordering techniques in parallel coordinates using crossing angles and mutual information.

\begin{figure}[h]
\begin{subfigure}{0.48\textwidth}
\includegraphics[width=\textwidth, angle=0]{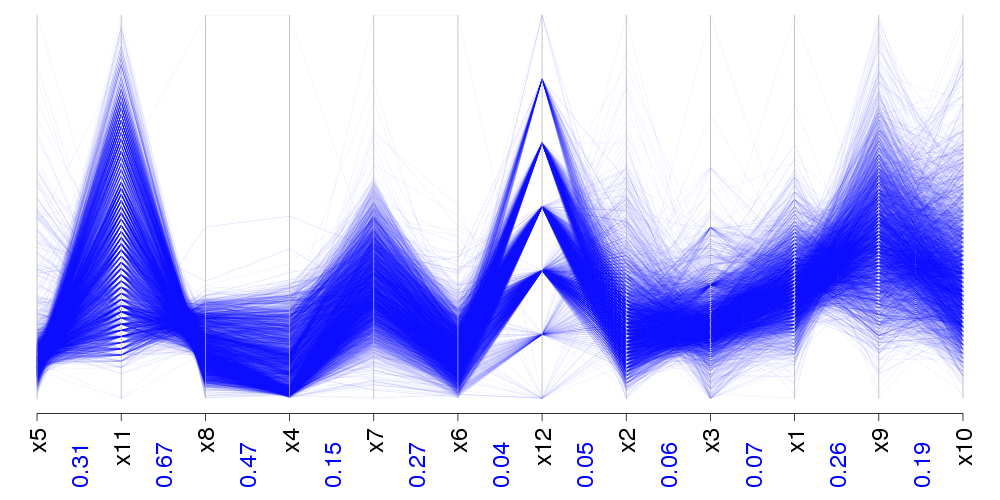}
\caption {~Order with CPLEX, $\sum_i \sum_j \GI(x_i, x_j)=2.53$.}

\end{subfigure} 
\begin{subfigure}{0.48\textwidth}
\includegraphics[width=\textwidth, angle=0]{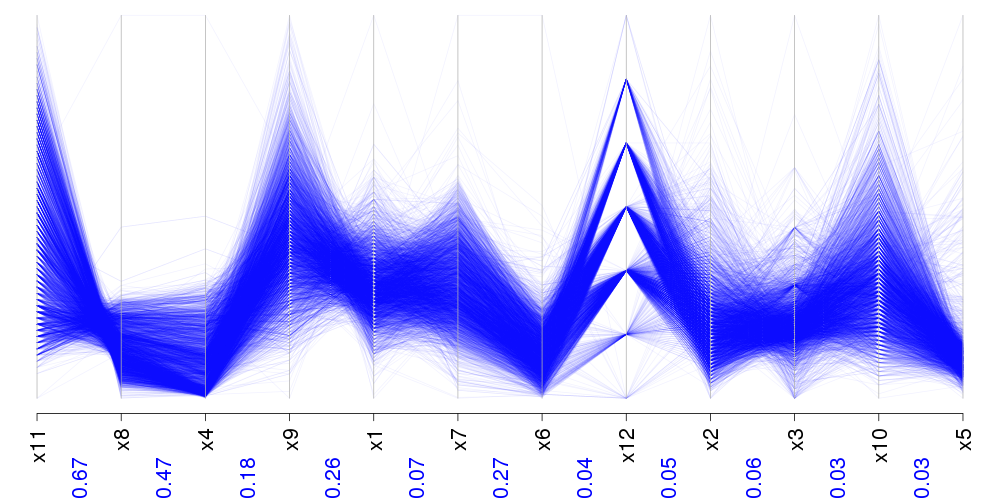}
\caption {~Order with greedy algorithm, $\sum_i \sum_j \GI(x_i, x_j)=2.13$}
\end{subfigure} 
\caption{Comparison between the order found with CPLEX (top panel) and the order with our Greedy algorithm (bottom panel). A transparency is applied to the polylines. The blue values between a pair of coordinates are $\GI(x_i,x_j)$.}
\label{ord:cplex:greedy}
\end{figure}

First, the optimal order using mutual information of the CPLEX optimizer is compared to the solution found by the greedy algorithm. The optimization problem is solved using IBM ILOG CPLEX Optimization Studio 12.7.1 a 2.20 GHz Intel core i7-2702MQ processor with 16.00 Go RAM. On our device it takes around $17$ seconds, while it takes only $1$ second using our greedy algorithm. The optimal solution given by CPLEX is a circle-like neighborhood matrix. To transform this neighborhood matrix it into a list, the circle is cut at the pair with the minimum mutual information. 
Figure~\ref{ord:cplex:greedy} presents a comparison between the order given by CPLEX and the order given with our greedy algorithm. Many pairs of adjacent attributes appear in both panels ($x_8,x_{11}$), ($x_8,x_4$), ($x_7,x_6$), ($x_6,x_{12}$), ($x_{12},x_2$), ($x_2,x_3$), and  ($x_1,x_9$).

\begin{figure}[h!]
\begin{subfigure}{0.48\textwidth}
\includegraphics[width=\textwidth, angle=0]{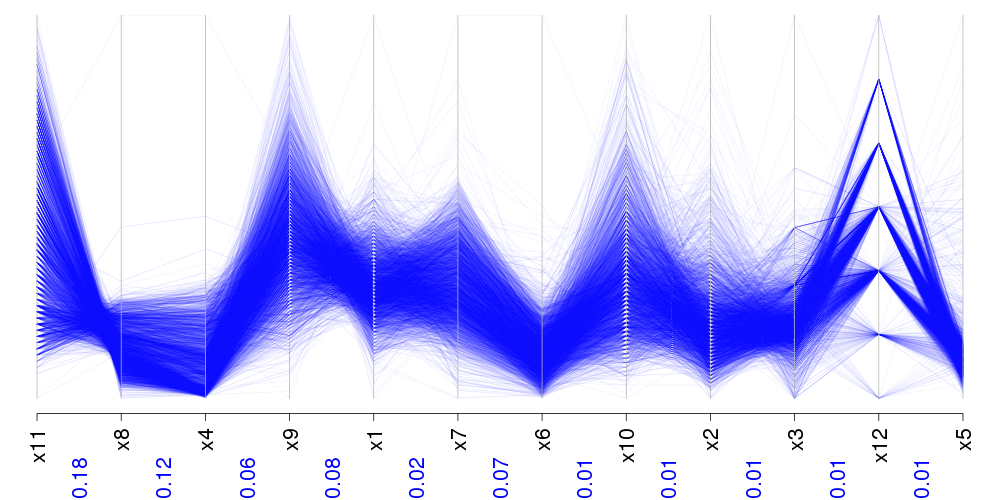}
\caption {~ Freeman-Tukey, $\sum_i \sum_j \GI(x_i,x_j)=0.58$}
\label{ord:tukey}
\end{subfigure} 
\begin{subfigure}{0.48\textwidth}
\includegraphics[width=\textwidth, angle=0]{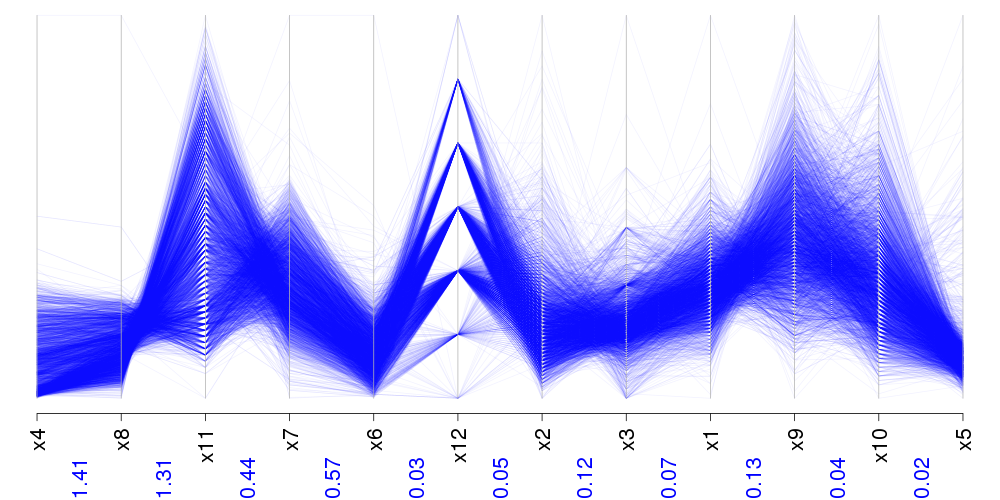}
\caption {~ Neyman, $\sum_i \sum_j \GI(x_i,x_j)=4.18$}
\end{subfigure} 
\begin{subfigure}{0.48\textwidth}
\includegraphics[width=\textwidth, angle=0]{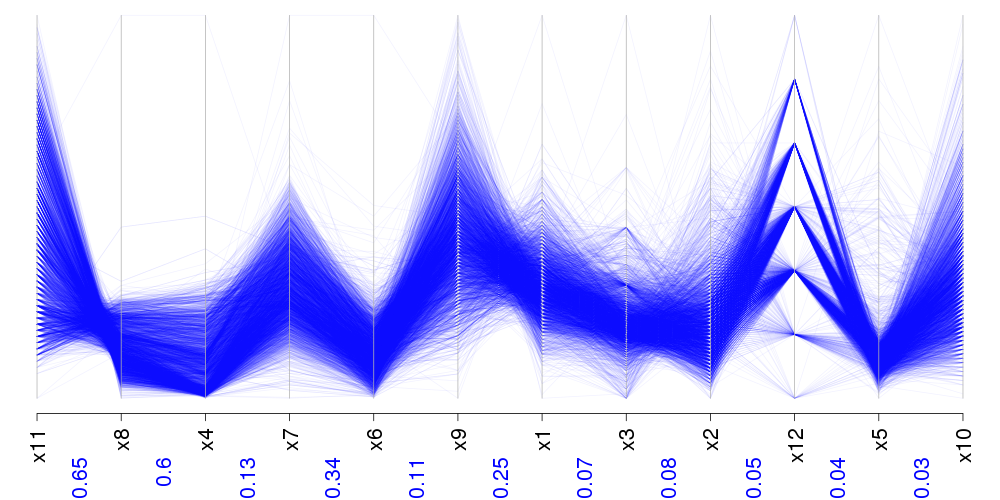}
\caption {~ Cressie-Read, $\sum_i \sum_j \GI(x_i,x_j)=2.39$}
\end{subfigure} 
\begin{subfigure}{0.48\textwidth}
\includegraphics[width=\textwidth, angle=0]{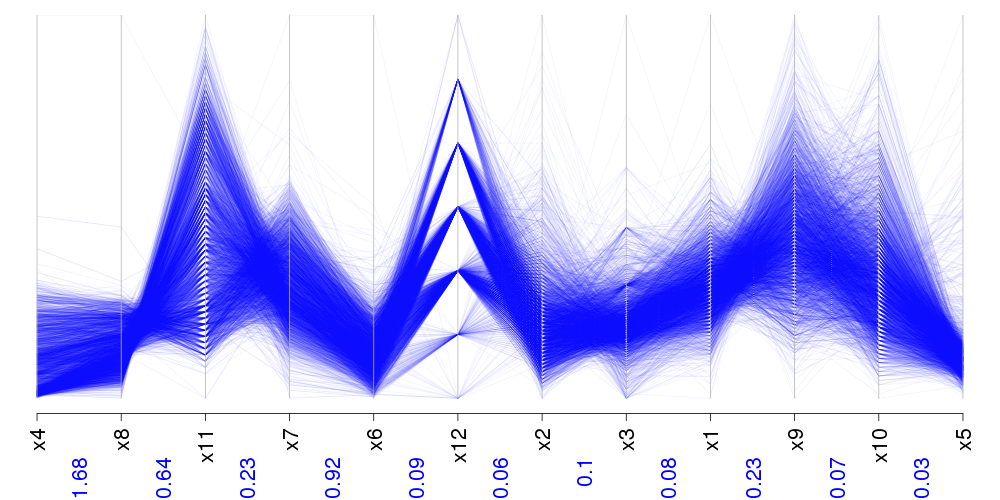}
\caption {~ Pearson chi-square, $\sum_i \sum_j \GI(x_i,x_j)=4.13$}
\label{ord:pearson}
\end{subfigure} 
\caption{Wine data reordered based on different statistics.A transparency is applied to the polylines. $\sum_i \sum_j \GI(x_i,x_j)$ is the sum of the values written between the pair of coordinates.}
\label{fig:wineorder1}
\end{figure}

The impact of changing the statistic by varying $G(.)$ is explored when the measuring concept ($F$ relative to $H$) is the dependence. Therefore, $F$ is the joint probability, $F(x_1,x_2)$, and $H$ is the product of marginal probability masses  $F(x_1) F(x_2)$.
The results are illustrated in Figure~\ref{fig:wineorder1}. The values between each $2$ adjacent attributes are the numerical values of the $\GI$ criterion. All the algorithms started with the highest information and tend to decrease.

The order changes as the statistic varies. For instance compare Figure~\ref{ord:tukey} with Figure~\ref{ord:pearson}. The first $3$ coordinates are ordered similarly by mutual information, Cressie-Read and Freeman-Tukey. Again, mutual information and Tukey-Freeman select the same $7$ attributes, but give a different order for the last $5$ attributes. In statistics literature it is known that the behaviour of the Neyman and the Pearson statistics are alike asymptotically. Here, Neyman statistic and Pearson statistic give exactly the same order. Tukey statistic, on the contrary, starts with a different attribute. However, Cressie-Read represents the dependence on attributes along with other statistics, for instance $(x_2,x_{12})$ and $(x_7,x_6)$ are present in the Pearson statistic.

Comparing the total information of each statistic, $\sum_i \sum_j \GI(x_i,x_j)$, shows that Neyman gives the highest value of $4.18$, followed by the Pearson with a similar total information $4.13$. Cressie-Read and mutual information come next, with total information around $2$, and Freeman statistic follows with a total information $0.58$. As the Neyman statistic and the Pearson statistic provide the highest total information between adjacent attributes, we suggest to use Neyman statistic or the Pearson statistic to reorder attributes for the wine dataset.

The order proposed by all criteria places the more dependent attributes first, and ending with nearly independent attributes. Changing the criterion, may change the order globally. However, many coordinates are placed in the neighborhood of one another overall.

\subsection{Genetic Dataset}
\label{subsec:genedata}
We applied the developed approach to \citet{Golub:1999un}. Golub dataset consists of $47$ patients with acute lymphoblastic leukemia (ALL) and $25$ patients with acute myeloid leukemia (AML). The observations have been assayed with Affymetrix Hgu6800 chips, resulting in $7129$ gene expressions (Affymetrix probes). The data was preprocessed, giving $2030$ attributes \citet{mcnicholas:2010:golub}.
This data is high-dimensional so, selecting the most informative attribute subset is crucial. Finding the genes that separates the data are more appealing than dependence in genetic application. Therefore, we apply the clustering statistic described earlier in \eqref{eq:onecomp}, by choosing $F$ to be a bivariate $k$-component Gaussian \eqref{eq:kcomp}, and $H$ to be a single component Gaussian.

\begin{figure*}[h!]
\begin{subfigure}{\textwidth}
\includegraphics[width=\textwidth, angle=0]{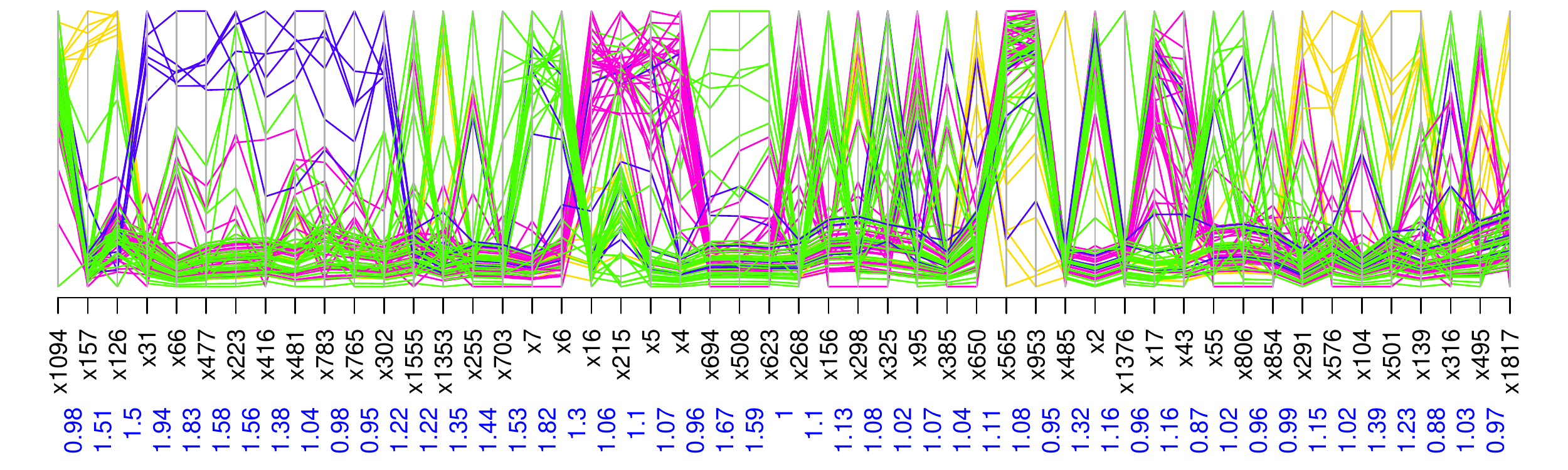}
\caption {~Golub data reordered based on separation criterion. The value between the adjacent axes is the general information adapted to measure data clustering, $\sum_i \sum_j \GI(x_i, x_j)=57$.}
\end{subfigure} 

\vspace{1in}

\begin{subfigure}{\textwidth}
\includegraphics[width=\textwidth, angle=0]{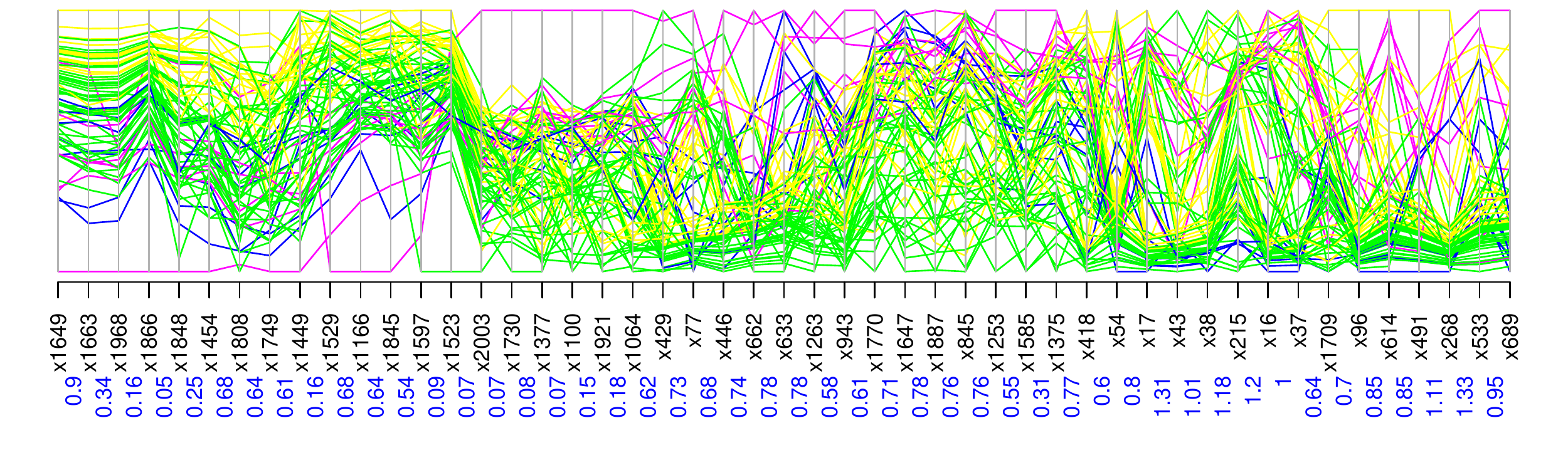}
\caption {~Golub data reordered based on Pearson correlation. The value between the adjacent axes is the general information adapted to measure data clustering $\sum_i \sum_j \GI(x_i, x_j)=30$.}
\label{fig:golub}
\end{subfigure} 
\end{figure*}

The visualized dimensions are those which maximize the criterion of the list. To find the appropriate order, we tried to run the optimization algorithm \eqref{eq:objective} using CPLEX, but it did not converge for $q=50$. Therefore, we only present the result of our greedy method. To improve the computational complexity of the greedy algorithm, the first attribute is selected to be the one with the highest univariate separation criterion. Fixing the first attribute avoids computation of general information criterion \eqref{eq:GI} for all pairs of attributes. This is a huge gain while data are high-dimensional.

When the number of clusters is not known, we suggest to use a large number of clusters to reordering, and then adjust the colors by combining small clusters for a better visualization.
This data  are clustered into $7$ clusters. Then, $3$ small clusters are re-grouped to visualize only $4$ groups. Figure~\ref{fig:golub} illustrates the results. The top panel shows clustered data, reordered based on the clustering criterion and the bottom panel shows clustered data, reordered based on Pearson correlation.

It is clear that for the purpose of cluster detection, clustering criterion highlights  the data separation more clearly. The sum of separation criterion is around $57$ for the order found based on the separation criterion and $30$ for the order based on Pearson correlation.  It is natural to expect that the total information for separation is higher when the attributes are reordered for this purpose. Not only the total information, but also the parallel coordinate graph clarifies the effect of choosing the right measure for the visualization purpose.

The result confirms that when the purpose of reordering is data clustering, or cluster detection as discussed in Section~\ref{sec:Intro}, then, $F$ and $H$ needs to defined in the direction of visualization purpose. 

\section{Conclusion}

\label{sec:conc}
We introduced a novel and a general framework for coordinate ordering. This framework is general enough to cover many existing ordering methods. Our approach uses a general information criterion to cover wide range of ordering measures. We also developed a computationally efficient ordering algorithm to cover high-dimensional data visualization.  Applying our approach on benchmark data shows the criterion and the statistic need to be chosen appropriately to achieve a visually meaningful coordinate order. Our framework is devised to build a coordinate ordering statistic that goes long with the visualization purpose.
This framework could be extended to propose a reordering measure that takes into account a output variable. This means that attributes would be reordered based on their contribution to the output variable.

\section*{Acknowledgement}
This research is supported by the Natural Sciences and Engineering Research Council of Canada (NSERC) Discovery grant.

\section*{Appendix}
Proof of Theorem~\ref{theo:asymp}. Without loss of generality assume contingency tables, i.e. $H(x_1, x_2)=F(x_1) F(x_2)$, proof for other nested $H$'s is similar. Let's change the notation to rows and columns $x_1 \in \{1,\ldots, I\}, x_2 \in \{1,\ldots,J\}$  $dF(x_1,x_2) = p_{ij},$ and $dH(x_1, x_2)=p_{i.}p_{.j},$ where $p_{i.}$ and $p_{.j}$ are the row and column marginal probabilities respectively. Therefore,  $$dF_n = \hat p_{ij}, \quad dH_n= \hat p_{i.} \hat p_{.j}.$$

Write the Taylor expansion of $G(u)$ about $1$ for $u={\pij \over \pi \pj}$ 
\begin{eqnarray*}
\hat\GI_n &=&{1\over G''(1)}\left\{ \sumi \sumj G(1) + {G'(1)\over 1!} (u-1) + {G''(1) \over 2! }(u-1)^2 + {G'''(1) \over 3! } (u-1)^3+ O_p(n^{-2})\right\}
\end{eqnarray*}
The first term is zero by definition, the second term vanishes asymptotically, the third term is the leading part, and the 4th term converges to zero with rate $o_p(n^{-1})$ if $G(u)$ is uniformly bounded on $u\in(1-\epsilon, 1+\epsilon).$
Therefore, 
\begin{eqnarray*}
\hat\GI_n &=& { {1 \over 2}  \sumi\sumj \left({\pij \over \pi \pj} -1\right)^2} + o_p(n^{-1})
\end{eqnarray*}
and 
\begin{eqnarray*}
{2n\hat\GI_n} &=& \chi^2_{\nu} + o_p(1), 
\end{eqnarray*}
with $\nu= \mathrm{dim}(F_n)- \mathrm{dim}(H_n).$
\bibliography{references}
\bibliographystyle{CUP}
\end{document}